%% file: single_dc_latency.tex
\documentclass[10pt, conference, letterpaper]{IEEEtran}
\usepackage{graphicx}
\newtheorem{theorem}{Theorem}

\newtheorem{lemma}{Lemma}

\newcommand{\be}{\begin{equation}}
\newcommand{\ee}{\end{equation}}
\newcommand{\bea}{\begin{eqnarray}}
\newcommand{\eea}{\end{eqnarray}}
\newcommand{\bw}{\begin{eqnarray*}}
\newcommand{\ew}{\end{eqnarray*}}

\newcommand{\D}{\displaystyle}

\usepackage{amssymb,amsmath}
\usepackage{amsmath}
\usepackage{latexsym}
\usepackage{graphicx}
\usepackage{graphics}
\usepackage{color}

\input{psfig.sty}
\usepackage{psfig}

\newenvironment{boxfig}[2]{\begin{figure}[#1]\fbox{\begin{minipage}{0.96\linewidth}
\vspace{0.2em}
\makebox[0.025\linewidth]{}
\begin{minipage}{0.95\linewidth}
{\small{#2}}
\end{minipage}
\end{minipage}}}{\end{figure}}

\begin{document}

\title{Differentiated latency in data center networks with erasure coded files through traffic engineering}
\author{Yu Xiang, Vaneet Aggarwal,   Yih-Farn R. Chen, and Tian Lan
}

\maketitle
\begin{abstract}
This paper proposes an algorithm to minimize weighted service latency for different classes of tenants (or service classes) in a data center network where erasure-coded files are stored on distributed disks/racks and access requests are scattered across the network. Due to limited bandwidth available at both top-of-the-rack and aggregation switches and tenants in different service classes need differentiated services, network bandwidth must be apportioned among different intra- and inter-rack data flows for different service classes in line with their traffic statistics.   We formulate this problem as weighted queuing and employ a class of probabilistic request scheduling policies to derive a closed-form upper-bound of service latency for erasure-coded storage with arbitrary file access patterns and service time distributions. The result enables us to propose a joint weighted latency (over different service classes) optimization over three entangled ``control knobs": the bandwidth allocation at top-of-the-rack and aggregation switches for different service classes, dynamic scheduling of file requests, and the placement of encoded file chunks (i.e., data locality). The joint optimization is shown to be a mixed-integer problem. We develop an iterative algorithm which decouples and solves the joint optimization as 3 sub-problems, which are either convex or solvable via bipartite matching in polynomial time. The proposed algorithm is prototyped in an open-source, distributed file system, {\em Tahoe}, and evaluated on a cloud testbed with 16 separate physical hosts in an Openstack cluster using 48-port Cisco Catalyst switches. Experiments validate our theoretical latency analysis and show significant latency reduction for diverse file access patterns. The results provide valuable insights on designing low-latency data center networks with erasure coded storage.

\end{abstract}

\input{Introv2}
\input{model}
\input{anal_v2}
\input{Opti}

\input{tahoe}

\input{Implep}

%\vspace{-.1in}

\section{Conclusions}

This paper considers an optimized erasure-coded single data center storage solution with differentiated services. The weighted latency of all file requests  is optimized over the placement and access of the erasure-coded contents of the file, and the bandwidth reservation at the switches for different service classes. The proposed algorithm shows significant reduction in average latency with the knowledge of the file-access patterns. Thus, this paper demonstrates that knowing the access probabilities of different files from different racks can lead to an optimized differentiated storage system solution, that works significantly better as compared to a storage solution which ignores the file access patterns.
%\vspace{-.1in}

%\vspace{-.1in}

%\input{Appendix}
\end{document}

%% file: Introv2.tex
\section{Introduction}

Data center storage is growing at an exponential speed and customers are increasingly demanding more flexible services exploiting the tradeoffs among the reliability, performance, and storage cost. Erasure coding has seen itself quickly emerged as a very promising technique to reduce the cost of storage while providing similar system reliability as replicated systems. The effect of coding on service latency to access files in an erasure-coded storage is drawing more attention these days. Google and Amazon have published that every 500 ms extra delay means a 1.2\% user loss \cite{Google}. Modern data centers often adopt a hierarchical architecture \cite{DC-Network}. As encoded file chunks are distributed across the system, accessing the files requires - in addition to data chunk placement and dynamic request scheduling - intelligent traffic engineering solutions to handle: 1) data transfers between racks that go through aggregation switches and 2) data transfers within a rack that go through Top-of-Rack (TOR) switches.

Quantifying exact latency for erasure-coded storage system is an open problem. Recently, there has been a number of attempts at finding latency bounds for an erasure-coded storage system \cite{MG1:12,Joshi:13,MDS-Queue,Yu-IFIP,CS14}. However, little effort has been made to address the impact of data ceter networking on service latency.

In this paper, we utilize the probabilistic scheduling policy developed in \cite{Yu-IFIP} and analyze service latency of erasure-coded storage with respect to data center topologies and traffic engineering. To the best of our knowledge, this is the first work that accounts for both erasure coding and data center networking in latency analysis. Our result provides a tight upper bound on mean service latency of erasure coded-storage and applies to general data center topology. It allows us to construct quantitative models and find solutions to a novel latency optimization leveraging both the placement of erasure coded chunks on different racks and the bandwidth reservations at different switches.

We consider a data center storage system with a hierarchical structure. Each rack has a TOR switch that is responsible for routing data flows between different disks and associated storage servers in the rack, while data transfers between different racks are managed by an aggregation switch that connects all TOR switches. Multiple client files are stored distributively using an $(n,k)$ erasure coding scheme, which allows each file to be reconstructed from any $k$-out-of-$n$ encoded chunks. We assume that file access requests may be generated from anywhere inside the data center, e.g., a virtual machine spun up by a client on any of the racks.  We  consider  an erasure-coded storage with multiple tenants and differentiated delay  demands.  While  customizing  elastic  service  latency for  the  tenants  is  undoubtedly  appealing  to  cloud  storage, it  also  comes  with  great  technical  challenges  and  calls  for a  new  framework  for  delivering,  quantifying,  and  optimizing differentiated service latency in general erasure coded storage.

Due to limited bandwidth available at both the TOR and aggregation switches, a simple First Come First Serve (FCFS) policy to schedule all file requests fails to deliver satisfactory latency performance, not only because the policy lacks the ability to adapt to varying traffic patterns, but also due to the need to apportion networking resources with respect to   heterogeneous service requirements. Thus,  we  study  a weighted queueing policy at top-of-the-rack and aggregation switches,  which partition  tenants  into  different  service  classes  based  on  their delay requirement and apply differentiated management policy to file requests generated by tenants in each service class.   We assume that the  file  requests  submitted  by  tenants in  each  class  are  buffered  and  processed  in  a  separate  first-come-first-serve queue at each switch. The service rate of  these  queues  are  tunable  with  the  constraints  that  they are  non-negative,  and  sum  to  at  most  the  switch/link capacity. Tuning these weights allows us to provide differentiated service rate to tenants in different classes, therefore assigning differentiated service latency to tenants.

In particular, a performance hit in erasure-coded storage systems comes when dealing with hot network switches. The latency of each file request is determined by the maximum delay in retrieving $k$-out-of-$n$ encoded chunks. Without proper coordination in processing each batch of chunk requests that jointly reconstructs a file, service latency is dominated by staggering chunk requests with the worst access latency (e.g., chunk requests processed by heavily congested switches), significantly increasing overall latency in the data center. To avoid this, bandwidth reservation can be made for routing traffic among racks \cite{YW14, NF12, SP11}. Thus, we apportion bandwidth at  TOR and aggregation switches so that each flow has its own FCFS queue for the data transfer. We jointly optimize bandwidth allocation and data locality (i.e., placement of encoded file chunks) to achieve service latency minimization.

For a given set of pair-wise bandwidth allocations among racks, and placement of different erasure-coded files in different racks, we first find an upper bound on mean latency of each file by applying the probabilistic scheduling policy proposed in \cite{Yu-IFIP}. Having constructed a quantitative latency model, we consider a joint optimization of average service latency (weighted by the rate of arrival of requests, and weight of the service class) over the placement of contents of each file, the bandwidth reservation between any pair of racks (at TOR and aggregation switches), and the scheduling probabilities.
This optimization is shown to be a mixed-integer optimization, especially due to the integer constraints for the content placement. To tackle this challenge, the latency minimization is decoupled into 3 sub-problems; two of which are proven to be convex. We propose an algorithm which iteratively minimizes service latency over the three engineering degrees of freedom with guaranteed convergence.

To validate our theoretical analysis and joint latency optimization for different tenants, we provide a prototype of the proposed algorithms in {\em Tahoe} \cite{Tahoe}, which is an open-source, distributed file system based on the {\em zfec} erasure coding library for fault tolerance. A Tahoe storage system consisting of 10 racks is deployed on hosts of virtual machines in an OpenStack-based data center environment, with each rack hosting 10 Tahoe storage servers. Each rack has a client node deployed to issue storage requests. The experimental results show that the proposed algorithm converges within a reasonable number of iterations. We further find that the service time distribution is nearly proportional to the bandwidth of the server, which is an assumption used in the latency analysis, and implementation results also show that our proposed approach significantly improved service latency in storage systems compared to native storage settings of the testbed.

%Finally, we see that our algorithm efficiently reduce latency with the weighted queues applied in our system model, and the results from the experiments are reasonably close to the given latency bounds for the model.

\begin{table}[th!]
\begin{center}
\tabcolsep=0.11cm
\begin{tabular}{cc}
\hline\hline \\[-0.7ex]
{\bf Symbol} & {\bf Meaning} \\ \hline \\[-0.7ex]
$N$& Number of racks, indexed by $i=1,\ldots,N$ \\[2ex]
$m$ & Number of storage servers in each rack  \\[2ex]
$R$ & Number of files in the system, indexed by $r=1,\ldots,R$ \\[2ex]
$(n,k)$ & Erasure code for storing files \\[2ex]
$D$ & Number of services classes \\[2ex]
$B$ & Total available bandwidth at the aggregate switch \\[2ex]
$b$ & Total available bandwidth at each top-of-the-rack switch  \\[2ex]
$b_{i,j}^{\rm eff}$ & Effective bandwidth for requests from node $i$ to node $j$ in the same rack \\[2ex]
$B_{i,j,d}^{\rm eff}$ & Effective bandwidth for servicing requests from rack $i$ to rack $j$ \\[2ex]
$w_{i,j,d}$ & Weight for apportioning top-of-rack switch bandwidth for service class $d$ \\[2ex]
$W_{i,j,d}$ & Weight for apportioning aggregate switch bandwidth  for service class $d$ \\[2ex]
$\lambda_{r_d}^i$ & Arrival rate of request for file $r$ of service class $d$ from rack $i$  \\[2ex]
%$\lambda_r$ & Aggregate arrival rate of request for file $r$ \\[2ex]
$\pi_{i,j,r_d}$ & Probability of routing rack-$i$ file-$r$ of service class $d $request to rack $j$  \\[2ex]
%$\Lambda_{i,j}$ & Aggregate arrival of requests from rack $i$ to rack $j$ \\[2ex]
$S_{r_d}$ & Set of racks for placing encoded chunks of file $r$ of service class $d$\\[2ex]
%$A_r$ & Set of racks for retrieval of chunks\\[2ex]
$N_{i,j}$ & Connection delay for service from rack $i$ to rack $j$ \\[2ex]
$Q_{i,j,d}$ & Queuing delay for service of class $d$ from rack $i$ to rack $j$ \\[2ex]
%$D_{i,j}$ & Total delay for service from rack $i$ to rack $j$ \\[2ex]
%$\mu, \sigma^2, \Gamma_t$ & Mean, variance and moments of standard service time $X$ \\[2ex]
$\bar{T}_{i,r_d}$ & Expected latency of a request of file $r$ from rack $i$ \\[2ex]
\hline\hline \\[0.5ex]
\end{tabular} \caption{Main notation.}
\end{center}
%\vspace{-0.3in}
\end{table}

%% file: model.tex
\section{System Model}

We consider a data center consisting of $N$ racks, denoted by $\mathcal{N}=\{i=1,2,\ldots,N\}$, each equipped with $m$ homogeneous servers that are available for data storage and running applications. There is a Top-of-Rack (TOR) switch connected to each rack to route intra-rack traffic and an aggregate switch that connects all TOR switches for routing inter-rack traffic. A set of files $\mathcal{R}=\{1,2,\ldots,R\}$ are distributedly stored among these $m*N$ servers and are accessed by client applications. Our goal is to minimize the overall file access latency of all applications to improve performance of the cloud system. 

This work focuses on a erasure-coded file storage systems, where files are erasure-coded and stored distributedly to enable space saving while allowing the same data durability as replication systems. More precisely, each file $r$ is split into $k$ fixed-size chunks and then encoded using an $(n,k)$ erasure code to generate $n$ chunks of the same size. The encoded chunks are assigned to and stored on $n$ out of $m*N$ distinct servers in the data center. While it is possible to place multiple chunks of the same file in the same rack\footnote{Using the techniques in \cite{Yu-IFIP}, our results in this paper can be easily extended to the case where multiple chunks of the same file are placed on the same rack.}, we assume that these $n$ servers to store file $r$ belong to distinct racks, so as to maximize the distribution of chunks across all racks, which in turn provides the highest reliability against independent rack failures. This is indeed a common practice adopted by file systems such as QFS\cite{QFS13}, an erasure-coded storage file system that is designed to replace HDFS for Map/Reduce processing.

For each file $r$, we use $\mathcal{S}_r$ to denote the set of racks that host encoded chunks of file $r$, satisfying $\mathcal{S}_r\subseteq\mathcal{N}$ and $|\mathcal{S}_r|=n$. To access file $r$, an $(n,k)$ Maximum-Distance-Separable (MDS) erasure code allows the file to be reconstructed from any subset of $k$-out-of-$n$ chunks. Therefore, a file request generated by a client application must be routed to a subset of $k$ racks in $\mathcal{S}_r$ for successful processing. To minimize latency, the selection of these $k$ racks needs to be optimized for each file request on the fly for load-balancing among different racks, in order to avoid creating ``hot" switches that suffer from heavy congestion and result in high access latency. We refer to this routing and rack selection problem as the {\em request scheduling problem}.

In this paper, we assume that the tenants' files are divided into $D$ service classes, denoted by class $d=1,2,\cdots, D$. Files in different service classes have different sensitivity and latency requirement. There are series of requests generated by client applications to access the $R$ files of different classes. We model the arrival of requests for each file $r$ in class $d$ as a Poisson process. To emphasize on each file's service class,  we denote a file $r$ in class $d$ by file $r_d$. Let $\lambda_{i,r_d}$ be the rate of file $r_d$ requests that are generated by a client application running in rack $i$. It means that a file $r_d$ request can be generated from a client application in any of the $N$ racks with a probability $\frac{\lambda_{i,r_d}}{\sum_i\lambda_{i,r_d}}$. The overall request arrival for file $r_{d}$ is a superposition of Poisson process with aggregate rate $\lambda_{r_d}=\sum_i\lambda_{r_d}^i$. Since solving the optimal request scheduling for latency minimization is still an open problem for general erasure-coded storage systems \cite{MG1:12,Joshi:13,MDS-Queue,Yu-IFIP}, we employ the {\em probabilistic scheduling policy} proposed in \cite{Yu-IFIP} to derive an outer bound of service latency, which further enables a practical solution to the request scheduling problem. 

Upon the arrival of each request, a probabilistic scheduler selects $k$-out-of-$n$ racks in $\mathcal{S}_r$ ( which host the desired file chunks) according to some known probability and route the resulting traffic to the client application accordingly. It is shown that determining the probability distribution of each $k$-out-of-$n$ combination of racks is equivalent to solving the marginal probabilities for scheduling requests $\lambda_{i,r_d}$,
\begin{eqnarray}
& \pi_{i,j,r_d}=\mathbb{P}[j\in \mathcal{S}_i^r {\rm \ is \ selected} \ | \ k \ {\rm racks \ are \ selected }],  \label{eq:pi}
\end{eqnarray}
under constraints $\pi_{i,j,r_d}\in [0,1]$ and $\sum_j \pi_{i,j,r_d} = k$ \cite{Yu-IFIP}. Here $\pi_{i,j,r_d}$ represents the scheduling probability to route a class $d$ request for file $r$ from rack $j$ to rack $i$. The constraint $\sum_j \pi_{i,j,r_d} = k$ ensures that exactly distinct $k$ racks are selected to process each file request. It is easy to see that the chunk placement problem can now be rewritten with respect to $\pi_{i,j,r_d}$. For example, $\pi_{i,j,r_d}=0$ equivalently means that no chunk of file $r$ exists on rack $j$, i.e., $j\notin \mathcal{S}_i^r$, while a chunk is placed on rack $j$ only if $\pi_{i,j,r_d}>0$.

To accommodate network traffic generated by applications in different service classes, bandwidth available at ToR and Aggregation Switches must be apportioned among different flows in a coordinated fashion. We propose a weighted queuing model for bandwidth allocation and enable differentiated latency QoS for requests belonging to different classes. In particular, at each rack $j$, we buffer all incoming class $d$ requests generated by applications in rack $i$ in a local queue, denoted by $q_{i,j,d}$. Therefore, each rack $j$ manages $N*D$ independent queues, among which include $D$ queues that manages intra-rack traffic traveling through the TOR switch and $(N-1)*D$ queues that manages inter-rack traffic of class $d$ originated from other racks $i\neq j$.

Assume the total bi-direction bandwidth at the aggregate switch is $B$, which needs to be apportioned among the $N(N-1)*D$ queues for inter-rack traffic. Let $\{W_{i,j,d}, \ \forall i\neq j\}$ be a set of $N(N-1)*D$ non-negative weights satisfying $\sum_{i,j:i\neq j} W_{i,j,d}=1$. We assign to each queue $q_{i,j,d}$ a share of $B$ that is proportional to $W_{i,j,d}$, i.e., queue $q_{i,j,d}$ receives a dedicated service bandwidth $B_{i,j,d}^{\rm eff}$ on the aggregate switch, i.e.,
\begin{eqnarray}
B_{i,j,d}^{\rm eff}= B\cdot W_{i,j,d}, \ \forall i\neq j. \label{eq:bandwidth1} \\ \nonumber
\sum_{d=1}^DW_{i,j,d}=W_{i,j}
\end{eqnarray}
where $W_{i,j}$ is the total weights assigned for traffic from rack $j$ to rack $i$. Because inter-rack traffic traverses two ToR switches, the same amount of bandwidth has to be reserved on the TOR switches of both racks $i$ and $j$. Then, any remaining bandwidth on the TOR switches will be made available for intra-rack traffic routing. On rack $j$, the bandwidth assigned for intra-rack traffic is given by the total TOR bandwidth $b$ minus aggregate incoming and outgoing inter-rack traffic ~\footnote{Our model, as well as the subsequent analysis and algorithms, can be trivially modified depending on whether the TOR switch is non-blocking and/or duplex.}, i.e.,
\begin{eqnarray}
b_{i,j}^{\rm eff}= b-\sum_{d=1}^D(\sum_{l: l\neq i} W_{i,l,d}B - \sum_{l: l\neq i} W_{l,i,d}B), \ \forall i=j. \label{eq:bandwidth2}
\end{eqnarray}
For different classes of intra-rack requests, we assume that requests of the same class will have the same service bandwidth, regardless the origin and destination ranks $i$ and $j$. Then this total available bandwidth at the TOR switch $b_{i,j}^{\rm eff}$ will be apportioned among $d$ queues, proportional to weights $w_{i,j,d}$. Then the efficient bandwidth for each class of intra-rack requests would be:
\begin{eqnarray}
B_{i,j,d}^{\rm eff}=w_{i,j,d}b_{i,j}^{\rm eff}, \ \forall i=j. \label{eq:bandwidth2}
\end{eqnarray}
By optimizing $W_{i,j,d}$ and $w_{i,j,d}$, the weighted queuing model provides an effective allocation of data center bandwidth among different classes of data flows both within and across racks. Bandwidth under-utilization and congestion needs to be addressed by optimizing these weights, so that queues with heavy workload will receive more bandwidth and those with light workload will get less. In the mean time, different classes of requests should also be assigned different shares of resources according to their class levels and to jointly minimize overall latency in the system.

Our goal in this work is to quantify service latency under this model through a tight upper bound and to minimize average service latency for all traffic in the data center by solving an optimization problem over three dimensions: placement of encoded file chunks $\mathcal{S}_r$, scheduling probabilities $\pi_{i,j,r_d}$ for load-balancing, and allocation of system bandwidth through weights for inter-rack and intra-rack traffic from different classes $W_{i,j,d}$ and $w_{i,j,d}$. To the best of our knowledge, this is the first work jointly minimizing service latency of an erasure-coded system over all three dimensions.
%\begin{figure}[thbp]
%\begin{center}
%\includegraphics[scale=0.32]{model_1.pdf}
%\caption{System model}
%\label{fig:system}
%\end{center}
%\end{figure}

%Fig ~\ref{fig:system} shows our system model with an example file request generated by rack 1. Due to $(5,3)$ erasure coding of files, 3 chunks from rack 1, rack 2 and rack 3 are retrieved respectively. Inter-rack data flows from rack 2 and rack 3 travel through the TOR switches and then are routed via the aggregate switch, which assigns the requests to 2 weighted queues with bandwidth $B_{2,1}^{\rm eff} = B\cdot w_{2,1}$, $B_{3,1}^{\rm eff} =B\cdot w_{3,1}$. The data flows are received by the TOR switch at rack 1 and then go to the destination server. On the other hand, intra-rack data flow is routed via the TOR switch of rack 1 only with available bandwidth $B_{1,1}^{\rm eff} = b-\sum_{j: j\neq 1} w_{1,j}B - \sum_{j: j\neq 1} w_{j,1}B$.

%{\color{red} Yu: Please read the new text and update Fig 1 and the example above. } 

%% file: anal_v2.tex
\section{Analyzing Service Latency for Data Requests}

In this section, we will derive an outer bound on the service latency of a file of class $d$ for each client application running in the data center. Under the probabilistic scheduling policy, it is easy to see that requests for file $r$ chunk from rack $i$ to rack $j$ form a (decomposed) Poisson process with rate $\lambda_{i,r_d}\pi_{i,j,r_d}$. Thus, the aggregate arrival of requests from rack $i$ to rack $j$ becomes a (superpositioned) Poisson process with rate
\begin{eqnarray}
\Lambda_{i,j,d}=\sum_{r}\lambda_{i,r_d}\pi_{i,j,r_d}.
\end{eqnarray}
It implies that if we limit our focus to a single queue, it can be modeled as $N^2*d$ M/G/1 queues. Service time per chunk is determined by the allocation of bandwidth $B\cdot W_{i,j,d}$ to each queue $q_{i,j,d}$ handling inter-rack traffic or the allocation of bandwidth $b_j$ to $q_{j,j,d}$ for different classes handling intra-rack traffic. Then service latency for each file request can be obtained by the largest delay to retrieve $k$ probabilistically-selected chunks of the file \cite{Yu-IFIP}. However, due to dependency of different queues (since their chunk request arrivals are coupled), it poses another challenge for deriving the exact service latency for file requests.

For each chunk request, we consider two latency components: \emph{connection delay} ${\bf N}_{i,j}$ that includes the overhead to set up the network connection between client application in rack $i$ and storage server in rack $j$ and \emph{queuing delay} ${\bf Q}_{i,j,d}$ that measures the time a chunk request experiences at the bandwidth service queue $q_{i,j,d}$, i.e., the time to transfer a chunk with the allocated network bandwidth. Let $\bar{T}_{i,r_d}$ be the average service latency for accessing file $r$ of class $d$ by an application in rack $i$. Due to the use of $(n,k)$ erasure coding for distributed storage, each file request is mapped to a batch of $k$ parallel chunk requests. The chunk requests are scheduled to $k$ different racks $\mathcal{A}_i^r\subset \mathcal{S}_r$ that host the desired file chunks. A file request is completed if all $k=|\mathcal{A}_i^r|$ chunks are successfully retrieved, which means a mean service time:
\begin{eqnarray}
\bar{T}_{i,r_d}=\mathbb{E}[\mathbb{E}_{\mathcal{A}_i^r} [\max\limits_{j\in \mathcal{A}_i^r}({\bf N}_{i,j}+{\bf Q}_{i,j,d})]], \label{eqn:order_statistic}
\end{eqnarray}
where the expectation is taken over queuing dynamics in the mode and over a set $\mathcal{A}_i^r$ of randomly selected racks with probabilities $\pi_{i,j,r_d}$ for all $j$.

Mean latency in (\ref{eqn:order_statistic}) could have been easily calculated if (a) ${\bf N}_{i,j}+{\bf Q}_{i,j,d}$'s are independent and (2) rack selection $\mathcal{A}_i^r$ is fixed.  However, due to coupled arrivals of chunk requests, the queues at different switches are indeed correlated. We deal with this issue by employing a tight bound of highest order statistic and extend it to account for randomly selected racks  \cite{Yu-IFIP}. This approach allows us to obtain a closed-form upper bound of average latency using only first and second moments of ${\bf N}_{i,j}+{\bf Q}_{i,j,d}$, which are easier to analyze. The result is summarized in Lemma 1.

\begin{lemma} ({\em Bound for the random order statistic} \cite{Yu-IFIP}.) For arbitrary set of $z_{i,r_d}\in\mathbb{R}$, the average service time is bounded by
\begin{eqnarray}
 & \bar{T}_{i,r_d} & \le \left\{ z_{i,r_d}+\sum_{j\in \mathcal{S}_r} \frac{\pi_{i,j,r_d}}{2}  \left(\mathbb{E}[{\bf D}_{i,j,d}] -z_{i,r_d} \right) +\sum_{j\in \mathcal{S}_r} \right.  \label{eq:lemma1} \nonumber\\
& & \left. \frac{\pi_{i,j,r_d}}{2} \left[ \sqrt{(\mathbb{E}[{\bf D}_{i,j,d}]-z_{i,r_d})^2+{\rm Var}[{\bf D}_{i,j,d}]}\right] \right\}
\end{eqnarray}
where ${\bf D}_{i,j,d} = {\bf N}_{i,j}+{\bf Q}_{i,j,d}$ is the combined delay. The tightest bound is obtained by optimizing $z_{i,r_d}\in \mathbb{Z}$ and is tight in the sense that there exists a distribution of ${\bf D}_{i,j,d}$ to achieve the bound with exact equality.
\end{lemma}

We assume that connection delay ${\bf N}_{i,j}$ is independent of queuing delay ${\bf Q}_{i,j,d}$. With the proposed system model, the chunk requests arriving at each queue $q_{i,j,d}$ form a Poison process with rate $\Lambda_{i,j,d}=\sum_{r}\lambda_{i,r_d}\pi_{i,j,r_d}$. Therefore, each queue $q_{i,j,d}$ at the aggregate switch can be modeled as a M/G/1 queue that processes chunk requests in an FCFS manner. For intra-rack traffic the $d$ queues at the top-of-rack switch at each rack can also be modeled as M/G/1 queues since the intra-rack chunk requests forms a Poison process with rate $\Lambda_{j,j,d}=\sum_r\lambda_{j,r_d}\pi_{j,j,r_d}$.  Due to the fixed chunk size in our system, we denote ${\bf X}$ as the standard (random) service time per chunk when bandwidth $B$ is available. We assume that the service time is inversely proportional to the bandwidth allocated to $q_{i,j,d}$, i.e., $B_{i,j,d}^{\rm eff}$.
%for queues $i \neq j$ handling inter-rack traffic and $b_j$ for queues $i=j$ handling intra-rack traffic. When a chunk request is processed on queue $q(i,j)$, 
We obtain the distribution of actual service time ${\bf X}_{i,j,d}$ considering bandwidth allocation:
\begin{eqnarray}
%{\bf X}_{i,j} \sim {\bf X}/w_{i,j} \ \ {\rm for} \ j\neq i  \\  
{\bf X}_{i,j,d} \sim {\bf X}\cdot B/B_{i,j,d}^{\rm eff}, \ \forall i,j
%\ \ {\rm for} \ j= i
\end{eqnarray}
%\begin{eqnarray}
%{\bf X}_{i,j} \sim {\bf X}\cdot B/B_{i,j}^{\rm eff} \ \ {\rm \forall} \ i,j
%\end{eqnarray}
%where $B_{i,j}^{\rm eff} $ can be found from Equation (\ref{eq:bandwidth1}) for $i\neq j$ and (\ref{eq:bandwidth2}) for $i\eq j$.\\

With the service time distributions above, we can derive the mean and variance of queuing delay ${\bf Q}_{i,j,d}$ using Pollaczek-Khinchine formula. Let $\mu=\mathbb{E}[{\bf X}]$, $\sigma^2={\rm Var}[{\bf X}]$, and $\Gamma_{t}=\mathbb{E}[{\bf X}^t]$, be the mean, variance, $t^\text{th}$ order moment of ${\bf X}$, $\eta_{i,j}$ and $\xi_{i,j}^2$ are mean and variance for connection delay ${\bf N}_{i,j}$. These statistics can be readily available from existing work on network delay \cite{AY11,WK} and file-size distribution \cite{D11,PT12}.

\begin{lemma}
The mean and variance of combined delay ${\bf D}_{i,j}$ for any $i,j$ is given by
\begin{eqnarray}
& \mathbb{E}[{\bf D}_{i,j,d}] & = \eta_{i,j}+ \frac{\Lambda_{i,j,d}\Gamma_{2}B^2}{2B_{i,j,d}^{eff}(B_{i,j,d}^{\rm eff}-\Lambda_{i,j,d}\mu B)} \\ \label{mean_Q}
& {\rm Var}[{\bf D}_{i,j,d}] & = \xi_{i,j}^2+\frac{\Lambda_{i,j,d}\Gamma_{3} B^3}{3(B_{i,j,d}^{\rm eff})^2(B_{i,j,d}^{\rm eff}-\Lambda_{i,j,d}\mu B)}+\nonumber \\
& & \frac{\Lambda_{i,j,d}(\Gamma_{2})^2 B^4}{4(B_{i,j,d}^{\rm eff})^2(B_{i,j,d}^{\rm eff}-\Lambda_{i,j,d}\mu B)^2} \label{var_Q}
\end{eqnarray}
where $B_{i,j,d}^{\rm eff}$ is the effective bandwidth assigned to the queue of requests for class $d$ files from rack $j$ to rack $i$.
%, i.e., $B_{i,j}^{\rm eff}=w_{i,j}B$ for inter-rack traffic with $j\neq j$ and $B_{i,j}^{\rm eff}=b_j$ for intra-rack traffic with $i=j$.
\end{lemma}

Combining these results,  we derive an upper bound for average service latency $\bar{T}_{i,r_d}$ as a function of chunk placement $\mathcal{S}_r$, scheduling probability $\pi_{i,j,r_d}$, and bandwidth allocation ${B}^{\rm eff}_{i,j,d}$  (which is a function of the bandwidth weights $W_{i,j,d}$ and $w_{i,j,d}$ in (\ref{eq:bandwidth1}) and (\ref{eq:bandwidth2})). The main result of our latency analysis is summarized in the following theorem.

\begin{theorem} \label{th:bound}
For arbitrary set of $z_{i,r_d}\in\mathbb{R}$, the expected latency $\bar{T}_{i,r_d}$ of a request of file $r$, requested from rack $i$ is upper bounded by
\begin{eqnarray}
 \bar{T}_{i,r_d} = z_{i,r_d}+ \sum_{j\in S_r } [\frac{\pi_{i,j,r_d}}{2} \cdot f(z_{i,r_d}, \Lambda_{i,j,d} ,  {B}^{\rm  eff}_{i,j,d}) ], \label{eq:bound}
\end{eqnarray}
where function $f(z_{i,r_d}, \Lambda_{i,j,d} , {B}^{\rm eff}_{i,j})$ is an auxiliary function depending on aggregate rate $\Lambda_{i,j,d}$ and effective bandwidth ${B}^{\rm eff}_{i,j,d}$ of queue $q_{i,j,d}$, i.e.,
\vspace{-2mm}
\begin{eqnarray}
& & f(z_{i,r_d}, \Lambda_{i,j,d} ,  {B}^{\rm eff}_{i,j,d}) = H_{i,j,d}+\sqrt{H_{i,j,d}^2+G_{i,j,d}} \label{eq:f1} \\
& & H_{i,j,d}=\eta_{i,j}+\frac{\Lambda_{i,j,d}\Gamma_{2} B^2}{2{B}^{\rm eff}_{i,j,d}({B}^{\rm eff}_{i,j,d}-\Lambda_{i,j,d} \mu B)}-z_{i,r_d} \label{eq:f2} \\
& & G_{i,j,d} = \xi_{i,j}^2+\frac{\Lambda_{i,j,d}\Gamma_{3} B^3}{3(B_{i,j,d}^{\rm eff})^2(B_{i,j,d}^{\rm eff}-\Lambda_{i,j,d}\mu B)} \nonumber \\
& & \ \ \ \ \ \ \ \ \ \ \ \ \ +\frac{\Lambda_{i,j,d}\Gamma_{2}^2 B^4}{4(B_{i,j,d}^{\rm eff})^2(B_{i,j,d}^{\rm eff}-\Lambda_{i,j,d} \mu B)^2} \label{eq:f3}
\end{eqnarray}
\end{theorem}

%&  f(z, \Lambda_{i,j} , w_{i,j}) & = \frac{\Lambda_{ij}\Gamma_2}{2w_{i,j}(w_{i,j}-\Lambda_{i,j}\mu)}-z+ \nonumber \\
%& & \sqrt{(\frac{\Lambda_{ij}\Gamma_2}{2w_{i,j}(w_{i,j}-\Lambda_{i,j}\mu)}-z)^2}+ \nonumber \\
%& &\overline{\frac{\Lambda_{i,j}\Gamma_3}{3w_{i,j}^2(w_{i,j}-\Lambda_{i,j}\mu)}+}\nonumber \\
%& &\overline{\frac{\Lambda_{i,j}\Gamma_2}{4w_{i,j}^2(w_{i,j}-\Lambda_{i,j}\mu)^2}}, \label{eq:bound1}

%For intra-rack traffic we have $f(z, \Lambda_{i,j} , w_{i,j})$  as:

%To simplify the expression, $f(z, \Lambda_{i,j} , w_{i,j})$ can also be described as:
%\begin{eqnarray}
%f(z, \Lambda_{i,j} , w_{i,j})=
%\end{eqnarray}
%where:
%$$H_{i,j}=\mathbb{E}[{\bf D}_{i,j}]-z$$
%$$G_{i,j}={\rm Var}[{\bf D}_{i,j}]$$
%For inter-rack queuing:
%$$\mathbb{E}[{\bf D}_{i,j}] = \eta_{i,j}+\frac{\Lambda_{i,j}\Gamma_2}{2w_{i,j}(w_{i,j}-\Lambda_{i,j}\mu)}$$
%$${\rm Var}[{\bf Q}_{i,j}] = \xi_{i,j}^2+\frac{\Lambda_{i,j}\Gamma_3}{3w_{i,j}^2(w_{i,j}-\Lambda_{i,j}\mu)}+\frac{\Lambda_{i,j}\Gamma_2^2}{4w_{i,j}^2(w_{i,j}-\Lambda_{i,j}\mu)^2}$$

%For intra-rack queuing:
%$$\mathbb{E}[{\bf Q}_{i,j}] = \eta_{i,j}+\frac{\Lambda_{i,j}\Gamma_2 B^2}{2B_{i,j}^{\rm eff}(B_{i,j}^{\rm eff}-\Lambda_{i,j}\mu B)} $$
%$${\rm Var}[{\bf Q}_{i,j}] =  \xi_{i,j}^2+\frac{\Lambda_{i,j}\Gamma_3 B^3}{3(B_{i,j}^{\rm eff})^2(B_{i,j}^{\rm eff}-\Lambda_{i,j}\mu B)}+\frac{\Lambda_{i,j}\Gamma_2^2 B^4}{4(B_{i,j}^{\rm eff})^2(B_{i,j}^{\rm eff}-\Lambda_{i,j}\mu B)^2}$$

%{\color{red} Yu, please complete the bound here as a function of $z, \Lambda_{i,j} , w_{i,j}$.}

%% file: Opti.tex
\section{Joint Latency Minimization in Cloud}

We consider a joint latency minimization problem over 3 design degrees of freedom: (1) placement of encoded file chunks $\{\mathcal{S}_r\}$ that determines datacenter traffic locality, (2) allocation of bandwidth at aggregation/TOR switches through weights for different classes $\{W_{i,j,d}\}$ and $\{w_{i,j,d}\}$ that affect chunk service time for different data flows, and (3) scheduling probabilities $\{\pi_{i,j,r_d}\}$ to optimize load-balancing under probabilistic scheduling policy and select racks/servers for retrieving files in different classes. Let $\lambda_{\rm all}= \sum_d\sum_i\lambda_{i,r_d}$ be the total file request rate in the datacenter. Let $\mathcal{F}_d$ be the set of files belonging to class $d$. The optimization objective is to minimize the mean service latency in the erasure-coded storage system, which is defined by
\begin{eqnarray}
& & \sum_{d=1}^D C_d\bar{T}_d, \ {\rm where} \ \bar{T}_d=\sum_{i=1}^N\sum_{r\in \mathcal{F}_d}\frac{\lambda_{i,r_d}}{\lambda_{\rm all}}T_{i,r_d}
\end{eqnarray}

Substitute $T_{i,r_d}$ in Equation (\ref{eq:bound}), we obtain $\bar{T}_d$ as:
\begin{eqnarray}
& \bar{T}_d = & \sum\limits_{i=1}^N \sum\limits_{r\in \mathcal{F}_d}\frac{\lambda_{i,r_d}}{\lambda_{all}}(z_{i,r_d} \nonumber \\
& & +\sum\limits_{j \in S_r}\frac{\pi_{i,j,r_d}}{2}f(z_{i,r_d},\Lambda_{i,j,d},B_{i,j,d}^{eff})) \nonumber \\
& \ \ \ \ \ = & \sum\limits_{i=1}^N\sum\limits_{r\in \mathcal{F}_d}[\frac{\lambda_{i,r_d}z_{i,r_d}}{\lambda_{all}}+ \nonumber \\
& &\sum\limits_{j\in S_r}\frac{\pi_{i,j,r_d}\lambda_{i,r_d}}{2\lambda_{all}}f(z_{i,r_d},\Lambda_{i,j,d},B_{i,j,d}^{eff})] \nonumber \\
& \ \ \ \ \ = & \sum\limits_{i=1}^N[\sum\limits_{r\in \mathcal{F}_d}\frac{\lambda_{i,r_d}z_{i,r_d}}{\lambda_{all}}+\nonumber \\
& & \sum\limits_{j=1}^N\frac{\Lambda_{i,j,d}}{2\lambda_{all}}f(z_{i,r_d},\Lambda_{i,j,d},B_{i,j,d}^{eff})]
\end{eqnarray}
 In the third step, we use the condition that $\pi_{i,j,r_d}=0$ for all $j\notin S_r$ to extend the limit of summation (because a rack not hosting a desired file chunk will never be scheduled) and exchange the order of summation. Step two follows from the fact that $\sum_{\text{r in class d}} \lambda_{i,r_d}\pi_{i,j,r_d}= \Lambda_{i,j,d}$.

We now define the Joint Latency and Weights Optimization (JLWO) problem as follows:
\begin{eqnarray}
& \d {\rm min} & \sum\limits_{d=1}^D C_d \bar{T}_d \label{eq:JLRM-SC} \\
& {\rm s.t.} & \bar{T}_d =\sum\limits_{i=1}^N[\sum\limits_{r\in \mathcal{F}_d}\frac{\lambda_{i,r_d}z_{i,r_d}}{\lambda_{all}}+\label{eq:JLRM-SC0}\nonumber \\
& & \sum\limits_{j=1}^N\frac{\Lambda_{i,j,d}}{2\lambda_{all}}f(z_{i,r_d},\Lambda_{i,j,d},B_{i,j,d}^{eff})] \\
& & \Lambda_{i,j,d} = \sum_{r\in \mathcal{F}_d} \lambda_{i,r_d}\pi_{i,j,r_d} \le \mu\frac{B_{i,j,d}^{\rm eff}}{B}, \nonumber \\
& & \forall i,j,d \label{eq:JLRM-SC1} \\
& &  \sum_{j=1}^N \pi_{i,j,r_d}= k \ {\rm and} \ \pi_{i,j,r_d} \in [0,1], \ \forall i,j,r_d \label{eq:JLRM-SC3}  \\
&  & |\mathcal{S}_r|=n \ {\rm and} \  \pi_{i,j,r_d}=0 \ \forall j\notin \mathcal{S}_i, \ \forall i,r_d  \label{eq:JLRM-SC4}  \\
&  & \sum\limits_{d=1}^D w_{i,j,d} =1.  \label{eq:JLRM-SC5}  \\
& & \sum\limits_{i=1}^N\sum\limits_{j\neq i}\sum_{d=1}^D W_{i,j,d}=1 \label{eq:JLRM-SC6}\\
& & B_{i,j,d}^{\rm eff} = W_{i,j,d}B\leq 1Gbps, \ \forall i\neq j \label{eq:JLRM-SC7} \\
& & B_{i,j,d}^{\rm eff} =w_{i,j,d}( b-\nonumber \\
& & \sum_{d=1}^D(\sum_{l: l\neq i} W_{i,l,d}B - \sum_{l: l\neq i} W_{l,i,d}B)) \leq C, \ \forall i= j \label{eq:JLRM-SC8} \\
%&  & |\mathcal{A}_i^r|=k \ {\rm and} \  \mathcal{A}_i^r\subseteq \mathcal{S}_i   \label{eq:JLRM-SC2}  \\
& {\rm var.} & z_{i,r_d}, \ \{\mathcal{S}_i^r\}, \ \{\pi_{i,j,r_d}\}, \ \{w_{i,j,d}\}, \ \{W_{i,j,d}\}. \nonumber
\end{eqnarray}
Here we minimize service latency derived in Theorem 1 over $z_{i,r_d}\in\mathbb{R}$ for each file access to get the tightest upper bound. Feasibility of Problem JLWO is ensured by (\ref{eq:JLRM-SC1}), which requires arrival rate to be no greater than chunk service rate received by each queue. Encoded chunks of each file are placed on a set $\mathcal{S}_i$ of servers in (\ref{eq:JLRM-SC4}), and each file request is mapped to $k$ chunk requests and processed by $k$ servers in $\mathcal{S}_i$ with probabilities $\pi_{i,j,r_d}$ in (\ref{eq:JLRM-SC3}). Finally, both weights $W_{i,j,d}$ (for inter-rack traffic) and $w_{i,j,d}$(for intra-rack traffic) should add up to 1 so there is no bandwidth left underutilized. Bandwidth assigned to each queue $B_{i,j,d}^{\rm eff}$ $\forall i,j$ is determined by our bandwidth allocation policy in (\ref{eq:bandwidth1}) and (\ref{eq:bandwidth2}) for intra- and inter-rack bandwidth allocations. Any effective flow bandwidth resulted from the bandwidth allocation have to be less than or equal to a port capacity $C$ allowed by the switches. For example, the Cisco switch used in our experiment, which has a port capacity constraint at $C=1Gbps$, which we will elaborate in Section \ref{implep}.

Problem JLWO is a mixed-integer optimization and hard to compute in general. In this work, we develop an iterative optimization algorithm that alternatively optimizes over the three dimension (chunk placement, request scheduling and inter/intra-rack weights assignment) of problem JLWO and solves each sub-problem repeatedly to generate a sequence of monotonically decreasing objective values. To introduce the proposed algorithm, we first recognize that Problem JLWO is convex in $\{\pi_{i,j,r_d}\}$, $\{w_{i,j,d}\}$, $\{W_{i,j,d}\}$ when other variables are fixed. The convexity will be shown by a sequence of lemmas as follows.

\begin{lemma}\label{th:lemma_3} ({\em Convexity of the scheduling sub-problem} \cite{Yu-IFIP}.) When $\{z_{i,r_d},W(w)_{i,j,d},S_r\}$ are fixed, Problem JLWO is a convex optimization over probabilities $\{\pi_{i,j,r_d}\}$.
\end{lemma}

\vspace{0.05in}
\noindent {\em Proof:} The proof is straightforward due to the convexity of ${\Lambda_{i,j,d}}f(z_{i,r_d}, \Lambda_{i,j,d} , B^{\rm eff}_{i,j,d})$ over $\Lambda_{i,j,d}$ (which is a linear combination of $\{\pi_{i,j,r_d}\}$) as shown in \cite{Yu-IFIP}, and the fact that all constraints are linear with respect to ${\pi_{i,j,r_d}}$. \hspace{1.7in}
$\square$
\vspace{0.05in}

\begin{lemma}\label{th:lemma_4} ({\em Convexity of the bandwidth allocation sub-problem.}) When $\{z_{i,r_d},\pi_{i,j,r_d},S_r, w_{i,j,d}\}$ are fixed, Problem JLWO is a convex optimization over inter-rack bandwidth allocation $\{W_{i,j,d}\}$. When $\{z_{i,r_d},\pi_{i,j,r_d},S_r, W_{i,j,d}\}$ are fixed, Problem JLWO is a convex optimization over intra-rack bandwidth allocation $\{w_{i,j,d}\}$.
\end{lemma}
\noindent {\em Proof:} Since all constraints in Problem JLWO are linear with respect to weights  $\{W_{i,j,d}\}$, we only need to show that the optimization objective $C_d\bar{T}_d=C_d\frac{\Lambda_{i,j,d}}{2\lambda_{all}}f(z_{i,r_d},\Lambda_{i,j,d},B_{i,j,d}^{eff})] $ is convex in $W_{i,j,d}$.Then we need to show that $f(z_{i,r_d}, \Lambda_{i,j,d} ,  {B}^{\rm eff}_{i,j,d})$ is convex in $\{W_{i,j,d}\}$ with other variables fixed. Notice that effective bandwidth ${B}^{\rm eff}_{i,j,d})$ is a linear function of the bandwidth allocation weights $\{W_{i,j}\}$ for both inter-rack traffic queues (\ref{eq:JLRM-SC6}) and intra-rack traffic queues (\ref{eq:JLRM-SC7}) when $\{w_{i,j,d}\}$ is fixed. Therefore, $f(z_{i,r_d}, \Lambda_{i,j,d} ,  {B}^{\rm eff}_{i,j,d})$ is convex in $\{W_{i,j,d}\}$ if it is convex in ${B}^{\rm eff}_{i,j,d})$. 

Toward this end, we consider $f(z_{i,r_d}, \Lambda_{i,j,d} ,  {B}^{\rm eff}_{i,j,d})= H_{i,j,d}+\sqrt{H_{i,j,d}^2+G_{i,j,d}} $ given in (\ref{eq:f1}), (\ref{eq:f2}) and (\ref{eq:f3}). 
Finally, to prove that $f= H_{i,j,d}+\sqrt{H_{i,j,d}^2+G_{i,j,d}} $ is convex in ${B}^{\rm eff}_{i,j,d}$, we have:
\begin{eqnarray}
\frac{\partial ^2 f}{\partial ({B}^{\rm eff}_{i,j,d})^2} &=& \frac{\partial^2 H_{i,j,d}}{\partial ({B}^{\rm eff}_{i,j,d})^2}+\frac{H_{i,j,d}\frac{\partial ^2 H_{i,j,d}}{\partial ({B}^{\rm eff}_{i,j,d})^2}+G_{i,j,d}\frac{\partial ^2 G_{i,j,d}}{\partial ({B}^{\rm eff}_{i,j,d})^2}}{(H_{i,j,d}^2+G_{i,j,d}^2)^{1/2}}+ \nonumber \\
& & \frac{(H_{i,j,d}\frac{d G_{i,j,d}}{d {B}^{\rm eff}_{i,j,d}}+G_{i,j,d}\frac{d H_{i,j,d}}{d {B}^{\rm eff}_{i,j,d}})^2}{(H_{i,j,d}^2+G_{i,j,d}^2)^{3/2}}
\end{eqnarray}
From where we can see that in order for $\frac{\partial ^2 f}{d ({B}^{\rm eff}_{i,j,d})^2}$ to be positive we only need $\frac{\partial ^2 H_{i,j,d}}{d ({B}^{\rm eff}_{i,j,d})^2}$ and $\frac{\partial ^2 G_{i,j,d}}{d ({B}^{\rm eff}_{i,j,d})^2}$ to be positive. 
We find the second order derivatives of $H_{i,j,d}$ with respect to ${B}^{\rm eff}_{i,j,d}$:
\begin{eqnarray}
 \frac{\partial ^2 H_{i,j,d}}{\partial ({B}^{\rm eff}_{i,j,d})^2}=\frac{\Lambda_{i,j,d}\Gamma_2(3({B}^{\rm eff}_{i,j,d})^2-3\Lambda_{i,j,d}\mu {B}^{\rm eff}_{i,j,d}-1)}{({B}^{\rm eff}_{i,j,d})^3({B}^{\rm eff}_{i,j,d}-\Lambda_{i,j,d}\mu)^3} 
\end{eqnarray}
which is positive as long as $1-\Lambda_{i,j,d}\mu/{B}^{\rm eff}_{i,j,d}>0$. This is indeed true because $\rho=\Lambda_{i,j,d}\mu/{B}^{\rm eff}_{i,j,d}<1$ in M/G/1 queues. Thus, $H_{i,j,d}$ is convex in ${B}^{\rm eff}_{i,j,d})$. Next, considering $G_{i,j}$ we have 
\begin{eqnarray}
\frac{\partial ^2 G_{i,j,d}}{\partial ({B}^{\rm eff}_{i,j,d})^2}=\frac{p({B}^{\rm eff}_{i,j,d})^3+q({B}^{\rm eff}_{i,j,d})^2+s{B}^{\rm eff}_{i,j,d}+t}{6({B}^{\rm eff}_{i,j,d})^4({B}^{\rm eff}_{i,j,d}-\Lambda_{i,j,d}\mu)^4}
\end{eqnarray}
where the auxiliary variables are given by
where we have:
$$p=24\Lambda_{i,j,d}\Gamma_3  $$
$$q=2\Lambda_{i,j,d}(15\Gamma_2^2 -28\Lambda_{i,j,d}\mu\Gamma_3)$$
$$s=2\Lambda_{i,j,d}^2\mu(22\Lambda_{i,j,d}\mu\Gamma_3-15\Gamma_2^2)$$
$$t=3\Lambda_{i,j,d}^3\mu^2(3\Gamma_2^2-4\Lambda_{i,j,d}\mu\Gamma_3)$$
which give out the solution for $p({B}^{\rm eff}_{i,j,d})^3+q({B}^{\rm eff}_{i,j,d})^2+s({B}^{\rm eff}_{i,j,d})+t$ as ${B}^{\rm eff}_{i,j,d}>\Lambda_{i,j,d}\mu$, which is equivalent to $1-\Lambda_{i,j,d}\mu/{B}^{\rm eff}_{i,j,d}>0$, which has been approved earlier. Thus $G_{i,j,d}$ is also convex in ${B}^{\rm eff}_{i,j,d}$.
 
Now that we can see $\frac{\partial ^2 f}{\partial ({B}^{\rm eff}_{i,j,d})^2}$ is positive and conclude that their composition $f(z_{i,r_d}, \Lambda_{i,j,d} ,  {B}^{\rm eff}_{i,j,d})$ is convex in ${B}^{\rm eff}_{i,j,d}$ and thus convex in $W_{i,j,d}$. So the objective is convex in $W_{i,j,d}$ as well.

Similarly, when we have $\{z_{i,r_d},\pi_{i,j,r_d},S_r, W_{i,j,d}\}$ as constants we will again have ${B}^{\rm eff}_{i,j,d}$ convex in $\{w_{i,j,d}\}$. And  the above proof for  $f(z_{i,r_d}, \Lambda_{i,j,d} ,  {B}^{\rm eff}_{i,j,d}$ is convex in ${B}^{\rm eff}_{i,j,d}$ still works, so that we can also conclude that $f(z_{i,r_d}, \Lambda_{i,j,d} ,  {B}^{\rm eff}_{i,j,d})$ is convex in intra-rack bandwidth allocation $\{w_{i,j,d}\}$ as well. 
This completes the proof. \hspace{2.0in} $\square$
\vspace{0.05in}

Next, we consider  the {\rm placement sub-problem} that minimizes average latency over $\{S_r\}$ for fixed $\{z_{i,r_d},\pi_{i,j}^r,w_{i,j}\}$. In this problem, for each file $r$ in class $d$ we permute the set of racks that contain each file $r$ to have a new placement $S'_r=\{\beta(j). \ \forall j \in S_r\}$ where $\beta(j) \in \mathcal{N}$ is a permutation. The new probability of accessing file $r$  from rack $\beta(j)$ when client is at rack $i$ becomes $\pi_{i,\beta{j},r_d}$. Our objective is to find such a permutation that minimizes the average service latency of class $d$ files, which can be solved via a matching problem between the set of scheduling probabilities $\{\pi_{i,j,r_d},\forall i\}$ and the racks, with respect to their load excluding the contribution of file $r$.
%For file $r$, we consider its scheduling probabilities $\{\pi_{i,j}^r,\forall r\}$ and divide them into $N$ disjoint groups, $\{\pi_{i1}^r,\forall i\}, \ \{\pi_{i2}^r,\forall i\}, \ \ldots, \ \{\pi_{iN}^r,\forall i\}$, each toward a different rack. Since scheduling probabilities are fixed in the placement sub-problem, the placement sub-problem is to match these $N$ groups of probabilities with destination racks $1, 2,\ldots,N$ in line with their existing load. Then, a rack $j$ is then selected to host a file-$r$ chunk if $\pi_{i,j}^r>0$ for some $i$. We notice that $S_r=n$ is guaranteed because any given scheduling probabilities that are feasible must already satisfy constraint (\ref{eq:JLRM-SC4}). 
Let $\Lambda_{i,j,d}^{-r} = \Lambda_{i,j,d} - \lambda_{i,r_d} \pi_{i,j,r_d}$ be the total request rate for class $d$ files between racks $i$ and $j$ excluding the contribution of file $r$. We define a complete bipartite graph $\mathcal{G}_r=(\mathcal{U},\mathcal{V},\mathcal{E})$ with disjoint vertex sets $\mathcal{U},\mathcal{V}$ of equal size $N$ and edge weights given by
\begin{eqnarray}
& & K_{i,j,d} = \sum_{i=1}^N [\sum_{\text {r is in class} d}\frac{\lambda_{i,r_d}z_{i,r_d}}{\lambda_{all}} \frac{\Lambda_{i,j,d}^{-r} + \lambda_{i,r_d} \pi_{i,j,r_d}}{\lambda_{\rm all}} \nonumber \\
& & f(z_{i,r_d}, \Lambda_{i,j,d}^{-r} + \lambda_{i,r_d} \pi_{i,j,r_d}, W(w)_{i,j,d})]. \label{eq:edge}
\end{eqnarray}
It is easy to see that a minimum-weight matching on $\mathcal{G}_r$ finds $\beta(j) \ \forall j$ to minimize
\begin{eqnarray}
& & \sum_{j=1}^N K_{i,\beta(j),d}  =  \sum_{i=1}^N\sum_{\text {r is in class} d}\frac{\lambda_{i,r_d}z_{i,r_d}}{\lambda_{all}} + \sum_{j=1}^N \sum_{i=1}^N   \nonumber  \\
& & \frac{\Lambda_{i,j,d}^{-r} + \lambda_{i.r_d} \pi_{i,\beta(j)},r_d }{\lambda_{\rm all}} \nonumber \\
& &  f(z_{i,r_d}, \Lambda_{i,j,d}^{-r} + \lambda_{i,r_d} \pi_{i,\beta(j)},r_d, W(w)_{i,j,d})
\end{eqnarray}
which is exactly the optimization objective of Problem JLWO if a chunk request of class $d$ is scheduled with probability $\pi_{i,\beta(j)},r_d$ to a rack with existing load $\Lambda_{i,j,d}^{-r}$.

\vspace{0.05in}

\begin{lemma}\label{th:lemma_5} ({\em Bipartite matching equivalence of the placement sub-problem.}) When $\{z_{i,r_d},\pi_{i,j,r_d},W(w)_{i,j,d}\}$ are fixed, the optimization of Problem JLWO over placement variables $S_r$ is equivalent to a balanced Bipartite matching problem of size $N$.
\end{lemma}
\noindent {\em Proof:} Based on the above statements, we can easily see that the placement sub-problem when other variables are all fixed is equivalent to find a minimum-weight matching on a bipartite graph whose of size $N$ whose edge weights $K_{i,j,d}$ are defined in \ref{eq:edge}, since the minimum weight matching is equal to the objective function of prolbem JLWO.

\vspace{0.05in}

Our proposed algorithm that solves Problem JLWO by iteratively solving the 3 sub-problems is summarized in Algorithm JLWO. It generates a sequence of monotonically decreasing objective values and therefore is guaranteed to converge to a stationary point. Notice that scheduling and bandwidth allocation sub-problems as well as the minimization over $z_{i,r_d}$ for each file are convex and can be efficiently computed by any off-the-shelf convex solvers, e.g., MOSEK \cite{MOSEK}. The placement sub-problem is a balanced bipartite matching that can be solved by Hungarian algorithm \cite{hung} in polynomial time.

\vspace{0.05in}

\begin{theorem}\label{th:thm2} The proposed algorithm generates a sequence of monotonically decreasing objective values and is guaranteed to converge.
\end{theorem}
\noindent {\em Proof:} Algorithm JLWO is working on three sub-problems iteratively and two of them (the scheduling and inter/intra-rack weights assignment) are convex optimization which will generates monotonically decreasing objective values. The placement sub-problem is an integer optimization and which can be solved by Hungarian algorithm, each iteration algorithm JLWO is solving the three sub-problems which converge individually, and since latency is bounded below, algorithm JLWO is guaranteed to converge to a fixed point of Problem JLWO.

\begin{boxfig}{th!}{\footnotesize
\begin{tabbing} \label{algo1}
xx\=xx\=xx\=xx\=xx\=xx\=\kill
{\bf Algorithm JLWO} : \\
\\
Initialize $t=0$, $\epsilon>0$. \\
Initialize feasible $\{z_{i,r_d}(0), \pi_{i,j,r_d}(0), S_r(0)\}$. \\
Initialize feasible $O^{(0)}$, $O^{(-1)}$ from (\ref{eq:JLRM-SC}) \\
User input $C_d$\\
\textbf{while} $O^{(t)} - O^{(t-1)}>\epsilon$ \\
\> // {\em Solve inter-rack bandwidth allocation for given} \\
\> //$\{z_{i,r_d}(t), \pi_{i,j,r_d}(t), w_{i,j,d}(t), S_r(t) \}$ \\
\> $W_{i,j,d}(t+1) = \D {\rm arg} \min_{W_{i,j,d}} $ (\ref{eq:JLRM-SC}) s.t. (\ref{eq:JLRM-SC0}), (\ref{eq:JLRM-SC1}), (\ref{eq:JLRM-SC6}), (\ref{eq:JLRM-SC7}).\\
\> // {\em Solve intra-rack bandwidth allocation for given}\\
\> // $\{z_{i,r_d}(t), \pi_{i,j,r_d}(t), W_{i,j,d}(t+1), S_r(t) \}$ \\
\> $w_{i,j,d}(t+1) = \D {\rm arg} \min_{w_{i,j,d}} $ (\ref{eq:JLRM-SC}) s.t. (\ref{eq:JLRM-SC0}), (\ref{eq:JLRM-SC1}), (\ref{eq:JLRM-SC5}),  (\ref{eq:JLRM-SC8}).\\
\> // {\em Solve scheduling for given} \\
\> //$\{z_{i,r_d}(t),S_r(t),w_{i,j,d}(t+1), W_{i,j,d}(t+1) \}$  \\
\> $\pi_{i,j,r_d}(t+1) = \D {\rm arg} \min_{\pi_{i,j,r_d}} $ (\ref{eq:JLRM-SC}) s.t. (\ref{eq:JLRM-SC1}), (\ref{eq:JLRM-SC3}). \\
\> // {\em Solve placement for given }\\
\>//$\{z_{i,r_d}(t),w_{i,j,d}(t+1),W_{i,j,d}(t+1),\pi_{i,j,r_d}(t+1) \}$ \\
\> \textbf{for} $d=1,\ldots,D$ \\
\> \> \textbf{for} $r$ in class $d$ \\
\> \> \> Calculate $\Lambda_{i,j,d}^{-r}$ using $\{\pi_{i,j,r_d}(t+1)\}$. \\
\> \> \> Calculate $K_{i,j,d}$ from (\ref{eq:edge}). \\
\> \> \> $(\beta(j) \forall j \in \mathcal{N})$=$Hungarian\_Algorithm( \{K_{i,j,d}\} )$. \\
\> \> \> Update $\pi_{i,\beta(j),r_d}(t+1) = \pi_{i,j,r_d}(t)$ $\forall i,j$. \\
\> \> \> Initialize $S_r(t+1)=\{\}$. \\
\> \> \> \textbf{for} $j=1,\ldots,N$ \\
\> \> \> \> \textbf{if} $\exists i$ s.t. $\pi_{i,j,r_d}(t+1) > 0$ \\
\> \> \> \> \> Update $S_r(t+1)=S_r(t+1) \cup \{j\}$. \\
\> \> \> \> \textbf{end if} \\
\> \> \> \textbf{end for} \\
\> \> \textbf{end for} {\color{white}{$\D \max_x$}} \\
\> \textbf{end for} \\
\> // {\em Solve $z_{i,r_d}$ given} \\
\> // $\{w_{i,j,d}(t+1), W_{i,j,d}(t+1), \pi_{i,j,r_d}(t+1), S_r(t+1) \}$ \\
\> $z_{i,r_d}(t+1) = \D {\rm arg} \min_{z_{i,r_d}\in\mathbb{R}} $. (\ref{eq:JLRM-SC}). \\
\> // {\em Update bound $T_{i,r_d}$ given} \\
\> // $\{w_{i,j,d}(t+1), W_{i,j,d}(t+1), \pi_{i,j,r_d}(t+1), S_r(t+1),$ \\
\> // $z_{i,r_d}(t+1)\}$ \\
\> Update objective value $O^{(t+1)}$=(\ref{eq:JLRM-SC}). \\
\> Update  $t=t+1$.\\
{\bf end while} \\
{\bf Output}: $\{\mathcal{S}_r(t), \pi_{i,j,r_d}(t), W_{i,j,d}(t), w_{i,j,d}(t), z_{i,r_d}(t) \}$ \\
\end{tabbing}
}
\end{boxfig}

%% file: tahoe.tex
\section{Implementation and Evaluation}
\label{implep}
\subsection{Tahoe Testbed}

To validate the weighted queuing model for $d$ different classes of files in our single data-center system model and evaluate the performance, we implemented the algorithms in {\em Tahoe} \cite{Tahoe}, which is an open-source, distributed file-system based on the {\em zfec} erasure coding library. It provides three special instances of a generic {\em node}: (a)  {\em Tahoe Introducer}: it keeps track of a collection of storage servers and clients and introduces them to each other.   (b) {\em Tahoe Storage Server}: it exposes attached storage to external clients and stores erasure-coded shares.  (c) {\em Tahoe Client}: it processes upload/download requests and connects to storage servers through a Web-based REST API and the Tahoe-LAFS (Least-Authority File System) storage protocol over SSL.

\begin{figure}[!thbp]
\begin{center}
\includegraphics[scale=0.32]{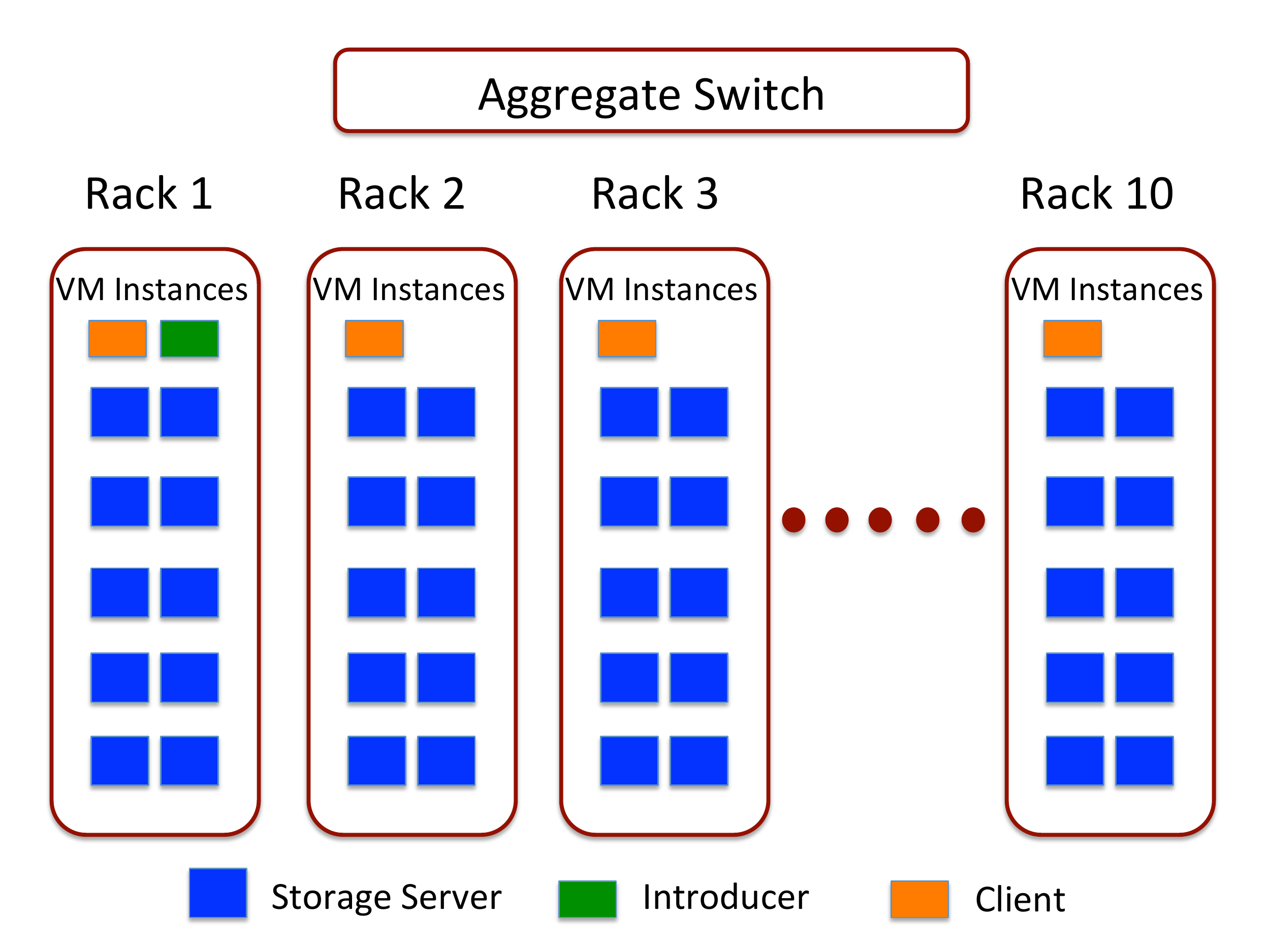}
\caption{Our Tahoe testbed with ten racks and distributed clients, each rack has 10 Tahoe storage servers.}
\label{fig:sl-testbed}
\end{center}
\vspace{-.2in}
\end{figure}
Our experiment is done on a Tahoe testbed that consists of 16 separate physical hosts in an Openstack cluster, 10 out of which have been used by our experiment. We simulated each host as a rack in the cluster. Each host has 11 VM instances, out of which 10 of them as running as Tahoe storage servers inside each rack and 1 one of them is running as a client node. Host 1 has 12 VM instances as there is one more instance running as Introducer in the Tahoe file system. We effectively simulated an Openstack cluster with 10 racks and each with 11 or 12 Tahoe storage servers on each rack. The cluster uses a Cisco Catalyst 4948 switch, which has 48 ports. Each port supports non-blocking, 1Gbps bandwidth in the full duplex mode. Our model is very general and can be applied to the case of $d$ classes of files. To simplify the implementation process, for experiments we consider only $d=2$ classes of files; each class has a different service rate, so the results will still be representative for cases of differentiated services in single data center. The 2-class weighted-queuing model is supposed to have $2*N(N-1)=180$ queues for inter-rack traffic at the aggregate switch; however, bandwidth reservation through ports of the switch is not possible since the Cisco switch does not support the OpenFlow protocol, so we made pairwise bandwidth reservations (using a bandwidth control tool from our Cloud QoS platform) between different Tahoe Clients and Tahoe storage servers. The Tahoe introducer node resides on rack 1 and each rack has a client node with multiple tahoe ports to simulate multiple clients to initiate requests of different classes of files coming from rack $i$. Our Tahoe testbed is shown in Fig ~\ref{fig:sl-testbed}.
%Each storage server has 1 VCPU, 2GB RAM and 20GB disk. Introducer also has 1 VCPU, 2GB RAM and 20GB disk. The client node in each rack has 2 VCPUs, 4GB RAM and 40GB disk. We use a larger client node so as to initiate a large number of requests through the client, running 10 threads at a time. All VMs have a 100GB volume attached for storing chunks(for storage servers and clients) and meta information(for introducer).

%% Robin:  I don't think you need to specify the detailed specs of the VM's

Tahoe is an erasure-coded distributed storage system with some unique properties that make it suitable for storage system experiments. In Tahoe, each file is encrypted, and is then broken into a set of segments, where each segment consists of $k$ blocks. Each segment is then erasure-coded to produce $n$ blocks (using an $(n,k)$ encoding scheme) and then distributed to (ideally) $n$ storage servers regardless of their server placement, whether in the same rack or not. The set of blocks on each storage server constitute a chunk. Thus, the file equivalently consists of $k$ chunks which are encoded into $n$ chunks and each chunk consists of multiple blocks\footnote{If there are not enough servers, Tahoe will store multiple chunks on one sever. Also, the term ``chunk'' we used in this paper is equivalent to the term ``share'' in Tahoe terminology. The number of blocks in each chunk is equivalent to the number of segments in each file.}.   For chunk placement, the Tahoe client randomly selects a set of available storage servers with enough storage space to store $n$ chunks. For server selection during file retrievals, the client first asks all known servers for the storage chunks they might have, again regardless of which racks the servers reside in. Once it knows where to find the needed k chunks (from among the fastest servers with a pseudo-random algorithm), it downloads at least the first segment from those servers. This means that it tends to download chunks from the ``fastest'' servers purely based on round-trip times (RTT). However, we consider RTT plus expected queuing delay and transfer delay as a measure of latency.

We had to make several changes in Tahoe in order to conduct our experiments. First, we need to have the number of racks $N \geq n$ in order to meet the system requirement that each rack can have at most one chunk of the original file.  In addition, since Tahoe has its own rules for chunk placement and request scheduling, while our experiment requires client-defined chunk placement in different racks,  and also with our server selection algorithms for placement and request scheduling in order to minimize joint latency for both classes of files,  we modified the upload and download modules in the Tahoe storage server and client to allow for customized and explicit server selection for both chunk placement and retrieval, which is specified in the configuration file that is read by the client when it starts. Finally, Tahoe performance suffers from its single-threaded design on the client side; we had to use multiple clients (multiple threads on one client node) in each rack with separate ports to improve parallelism and bandwidth usage during our experiments.

\subsection{Basic Experiment Setup}
We use (7,4) erasure code in the Tahoe testbed we introduced above throughout the experiments described in the implementation section.  The algorithm first calculates the optimal chunk placement through different racks for each class of files, which will be set up in the client configuration file for each write request. File retrieval request scheduling and weight assignment decisions of each class of files for inter-rack traffic also comes from Algorithm JLWO. The system calls a bandwidth reservation tool to reserve the assigned bandwidth $BW_{i,j,d}$ for each class of files based on optimal weights of each inter-rack pair; this is also done for intra-rack bandwidth reservation for each class $b_{i,j}^{eff}w_{i,j,d}$. Total bandwidth capacity at the aggregate switch is 96 Gbps, 48 ports with 1Gbps in each direction since we are simulating host machines as racks and VM's as storage servers. Intra-rack bandwidth as measured from \emph{iPerf} measurement is 792Mbps, disk read bandwidth for sequential workload is 354 Mbps, and write bandwidth is 127 Mbps. Requests are generated based on arrival rates at each rack and submitted from client nodes at all racks.

\subsection{Experiments and Evaluation}

\begin{figure*}[!htb]
\begin{minipage}{0.47\textwidth}
\begin{center}
\includegraphics[scale=.3]{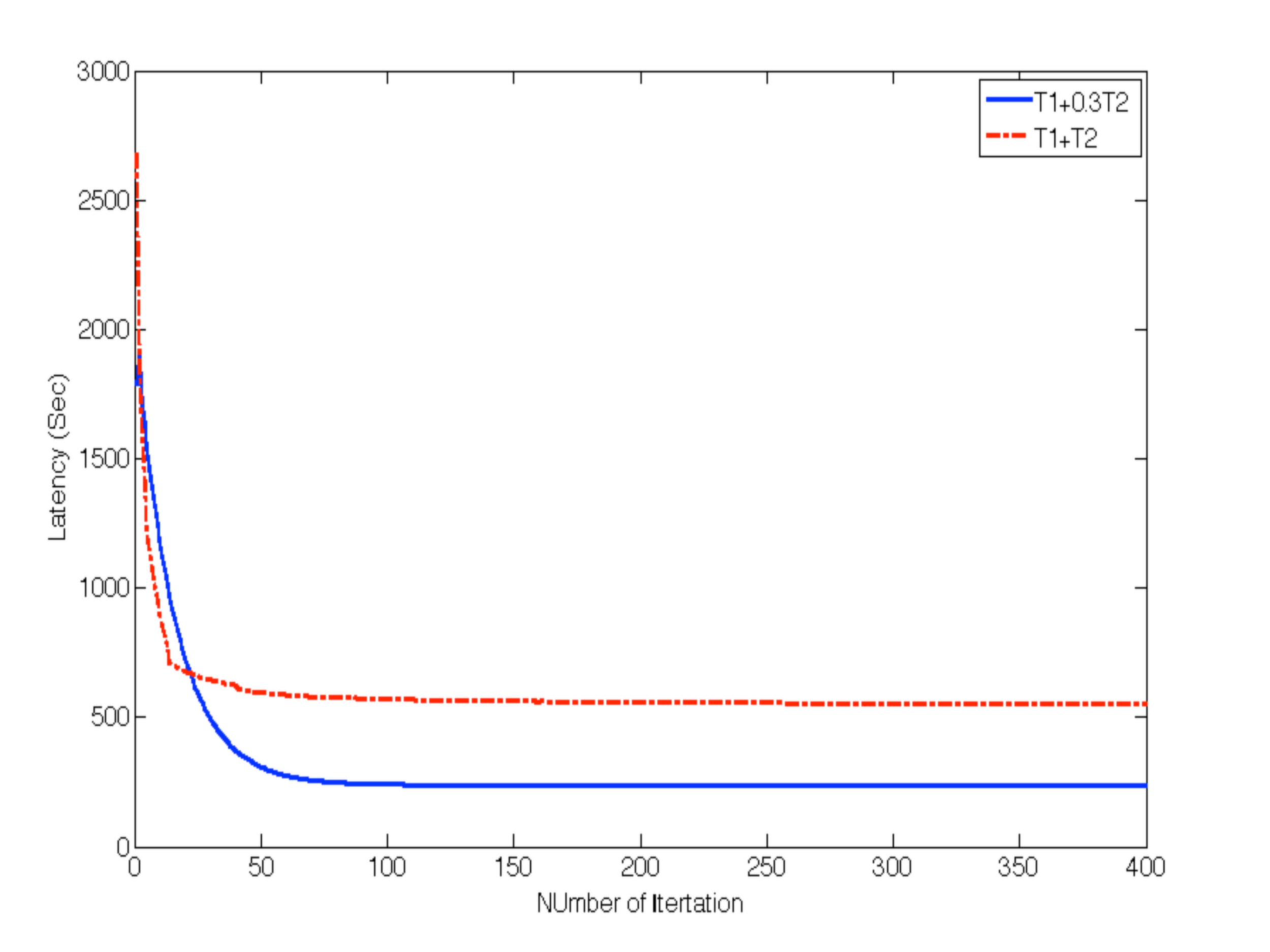}
\end{center}
\vspace{-0.2in}
\caption{\small Convergence of Algorithm JLWO with r=1000 requests for each of the two file classes for heterogeneous files from each rack on our 100-node testbed, with different weights on the second file class. Algorithm JLWO efficiently compute the solution in 140 iterations in both cases.}
\vspace{-0.1in}
\label{fig:AL}
\end{minipage}
\hspace{0.5in}
\begin{minipage}{0.45\textwidth}
\begin{center}
\includegraphics[scale=0.3]{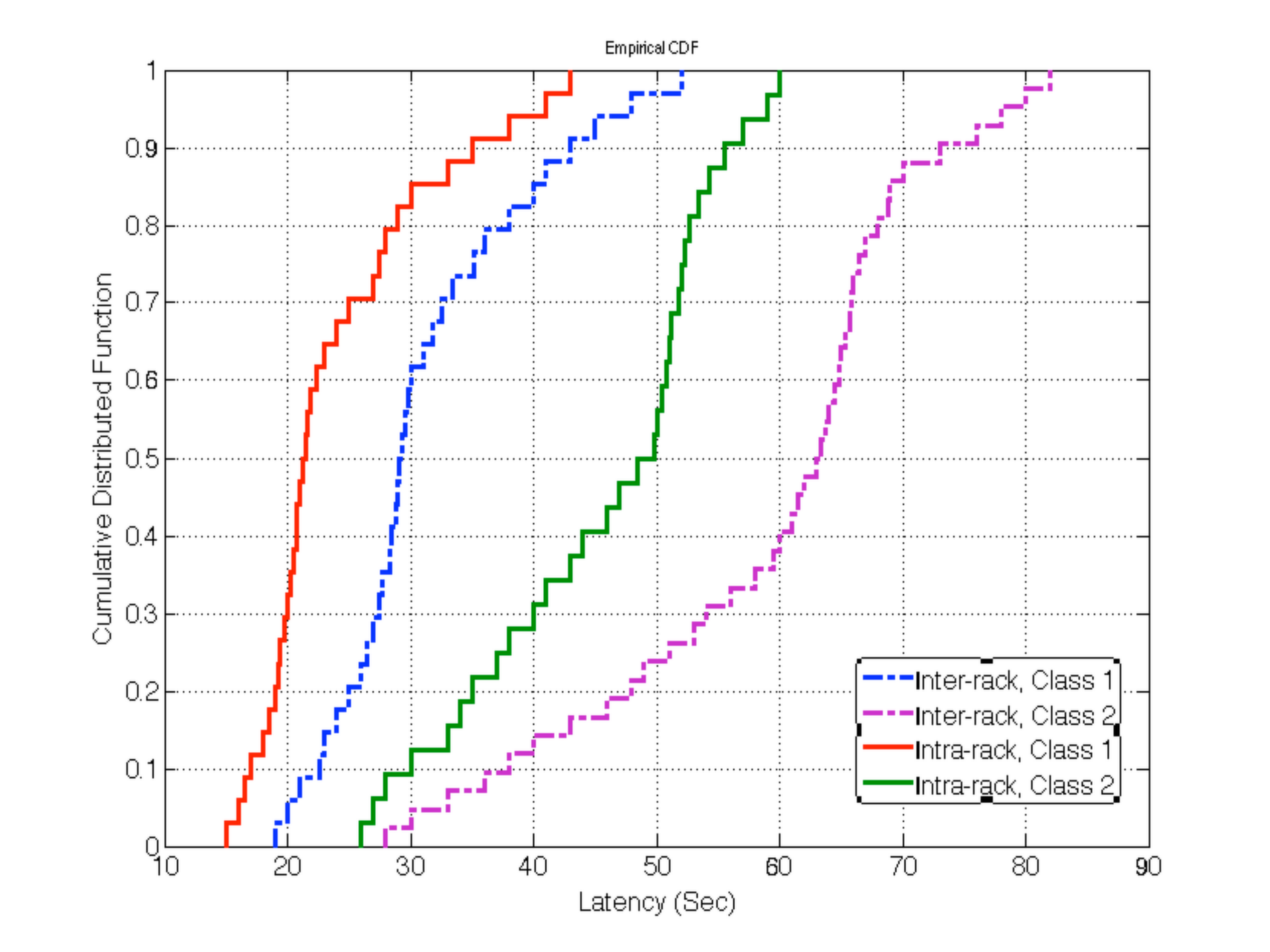}
\end{center}
\vspace{-0.2in}
\caption{\small Actual service time distribution of chunk retrieval through intra-rack and inter-rack traffic for both classes. Each of them has $1000$ files of size $100MB$ using erasure code (7,4) with the aggregate request arrival rate set to $\lambda_i=0.25$ /sec. Service time distributions indicated that chunk service time of the two classes are nearly proportional to the bandwidth reservation based on weight assignments for M/G/1 queues.}
\vspace{-0.1in}
\label{fig:cdf}
\end{minipage}
\end{figure*}

{\bf Convergence of Algorithm.}
We implemented Algorithm JLWO using MOSEK, a commercial optimization solver. With 10 racks and 10 simulated distributed storage servers on each rack, there are a total of 100 Tahoe servers in our testbed. Figure~\ref{fig:AL} demonstrates the convergence of our algorithms, which optimizes the latency of all file requests of the two classes coming from different racks for the weighted queuing model at the aggregate switch: chunk placement $\mathcal{S}_i$,  load balancing $\pi_{i,j}$ and bandwidth weights distribution at the aggregate switch and top-of-rack switch for each class $W_{i,j,d}$ and $w_{i,j,d}$. We fix $C_1=1$ and then set $C_2=1$ and $C_2=0.3$ to see the performance of algorithm JLWO when the importance of the second class varies. The JLCM algorithm, which has been applied as part of our JLWO algorithm, was proven to converge in Theorem 2 of \cite{Yu-IFIP}. In this paper, fisrt, we see the convergence of the proposed optimized queuing algorithms in Fig.~\ref{fig:AL}. By performing similar speedup techniques as in \cite{Yu-IFIP},  our algorithms efficiently solve the optimization problem with $r=1000$ files of each class at each of the 10 racks. Second, we can also see that as class 2 becomes as important as the class 1, the algorithm converges faster as it's actually solving the problem for one single class of files. It is observed that the normalized objective converges within 140 iterations for a tolerance $\epsilon=0.01$, where each iteration has an average run time of 1.19 sec, when running on an 8-core, 64-X86 machine, therefore the algorithm converges within 2.77 min on average from observation. To achieve dynamic file management, our optimization algorithm can be executed repeatedly upon file arrivals and departures.

\begin{figure}[!thbp]
\begin{center}
\includegraphics[scale=0.32]{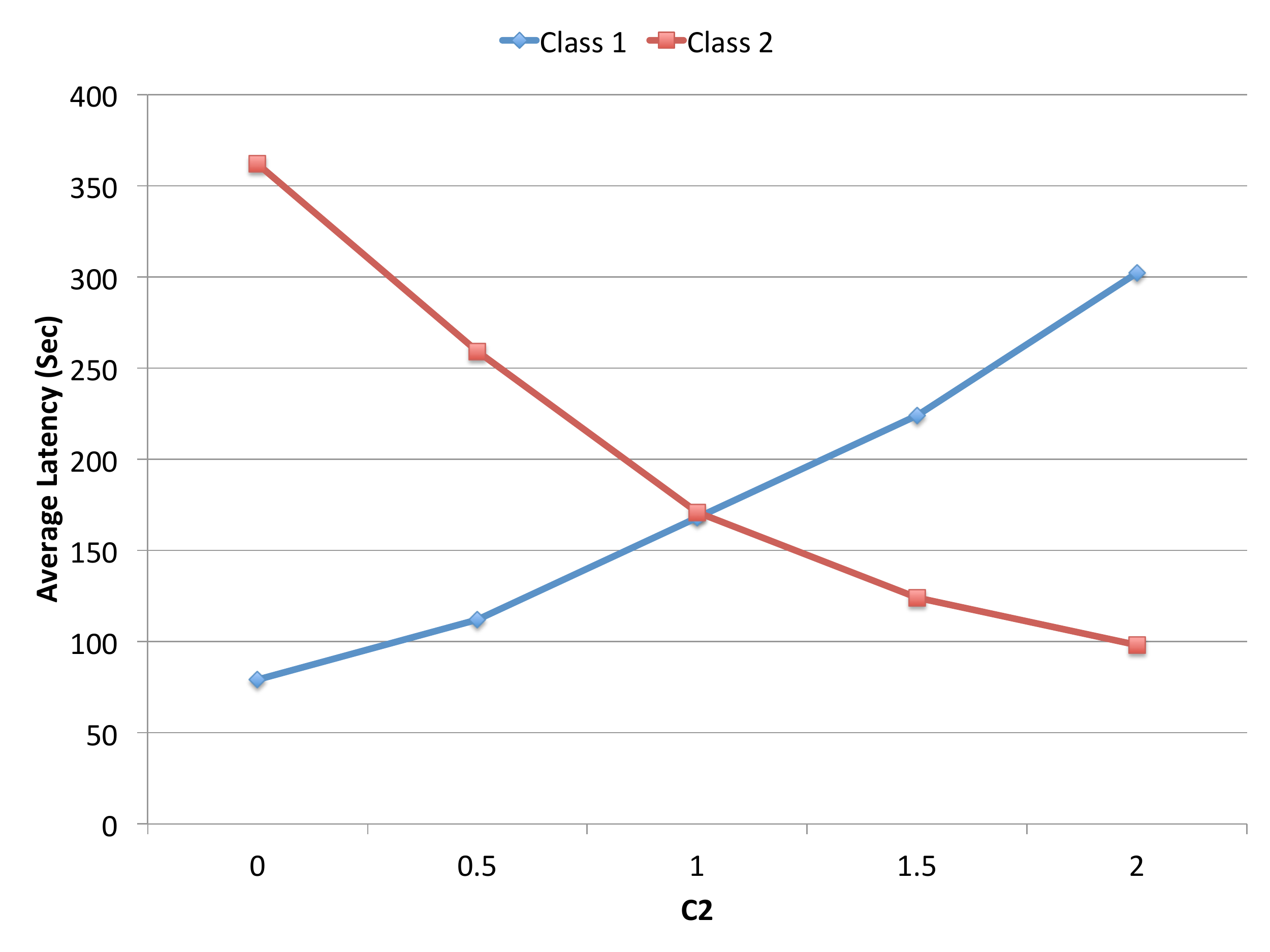}
\caption{file requests for different files of size 100MB, aggregate request arrival rate for both classes is 0.25/sec; varying C2 to validate our algorithms.}
\label{fig:sl-c2}
\end{center}
\vspace{-.2in}
\end{figure}

{\bf Validate Experiment Setup.} While our service delay bound applies to arbitrary distribution and works for systems hosting any number of files, we first run an experiment to understand actual service time distribution for both intra-rack and inter-rack retrieval in weighted queuing for the 2 classes of files on our testbed. We uploaded $r=1000$ files for each class of size $100MB$ file using a $(7,4)$ erasure code from the client at each rack based on the algorithm output of chunk placement from algorithm JLWO. And we have set $C_1=1$ and $C_2=0.4$. Inter-rack and intra-rack bandwidth was reserved based on the weights for each class output from the algorithm as well. We then initiated 1000 file retrieval requests for each class (each request for a unique file) from the clients distributed in the data center, using the algorithm output $\pi_{i,j,d}$ for retrieval request scheduling with the same erasure code. The experiment has 2000 file requests in total (from 10 racks), with an aggregate request arrival rate of 0.25/sec for clients at all racks and requests are evenly distributed across the racks. Based on the optimal sets for retrieving chunks of each file request provided by our algorithm, we get measurements of service time for both the inter-rack and intra-rack processes. The average inter-rack bandwidth over all racks for request of class 1 files is 674 Mbps and the intra/inter-rack bandwidth ratio is $889 Mbps/674 Mbps=1.318$.  The average inter-rack bandwidth over all racks for request of class 2 files is 684 Mbps and the intra/inter-rack bandwidth ratio is $632 Mbps/389 Mbps=1.625$.  Figure \ref{fig:cdf} depicts the Cumulative Distribution Function (CDF) of the chunk service time for both intra-rack and inter-rack traffic. We note that intra-rack requests for class 1 have a mean chunk service time of  22 sec and inter-rack chunk requests for class 1 have a mean chunk service time of 30 sec,  which is a ratio of 1.364 which is very close to the bandwidth ratio of 1.318. Also intra-rack requests for class 2 have a mean chunk service time of  50 sec and inter-rack chunk requests for class 2 have a mean chunk service time of 83 sec,  which is a ratio of 1.66 which is very close to the bandwidth ratio of 1.625.This means the chunk service time is nearly proportional to the bandwidth reservation on inter/intra-rack traffic for both classes of files.
%\begin{figure}[!thbp]
\begin{figure*}[!htb]
\begin{minipage}{0.47\textwidth}
\begin{center}
\includegraphics[scale=0.32]{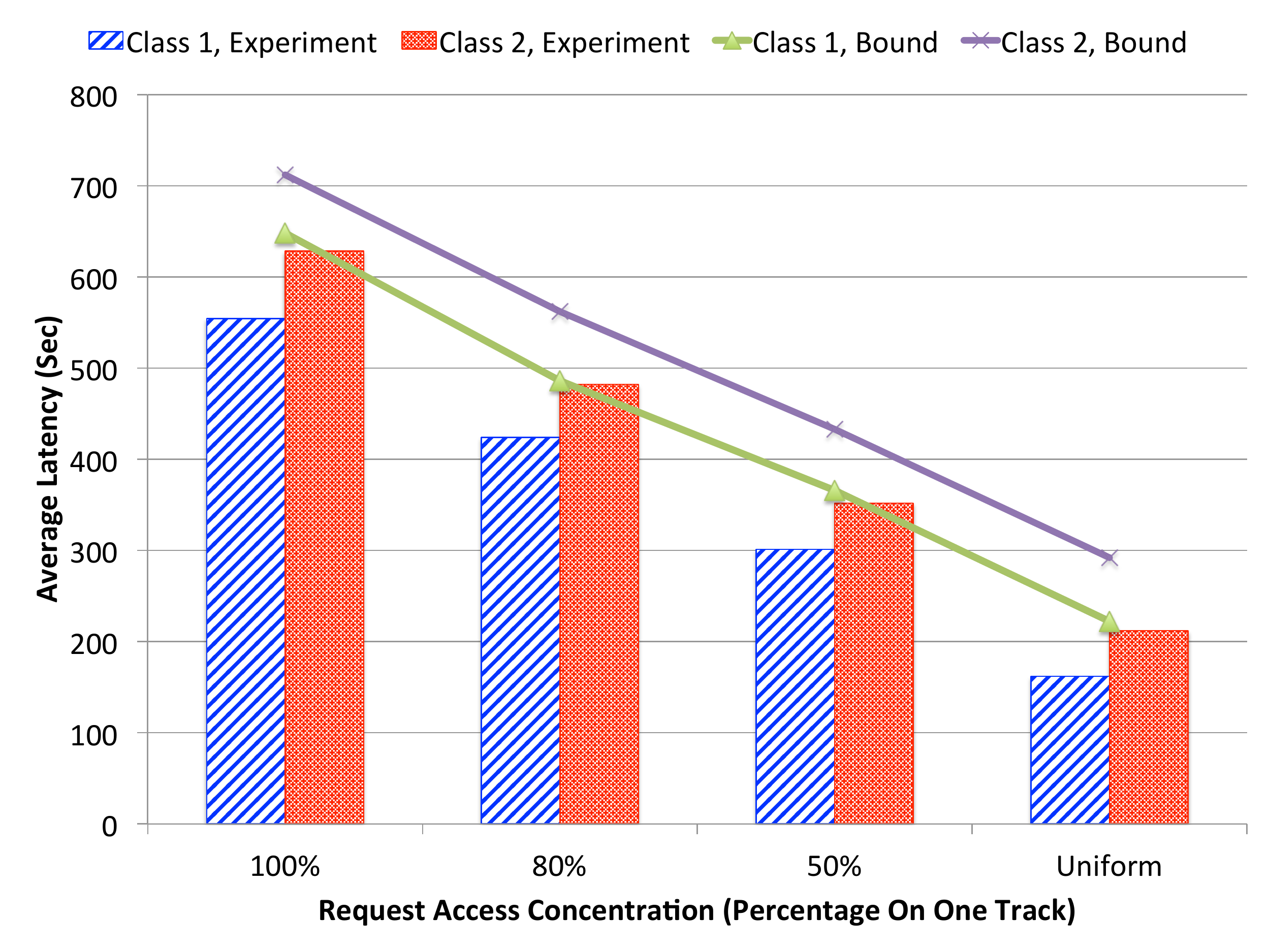}
\caption{\small Comparison of average latency with different access patterns. Experiment is set up for 1000 heterogeneous files for each class, there is one request for each file and 2000 file requests in total. Ratio for each class is 1:1. The figure shows the percentage that these 2000 requests are concentrated on the same rack. Aggregate arrival rate 0.25/sec, file size 200M. Latency improved significantly with weighted queuing.  Analytic bound for both classes tightly follows actual latency as well.}
\label{fig:access}
\end{center}
\vspace{-.2in}
\end{minipage}\hfill
\begin{minipage}{0.47\textwidth}
\begin{center}
\includegraphics[scale=.33]{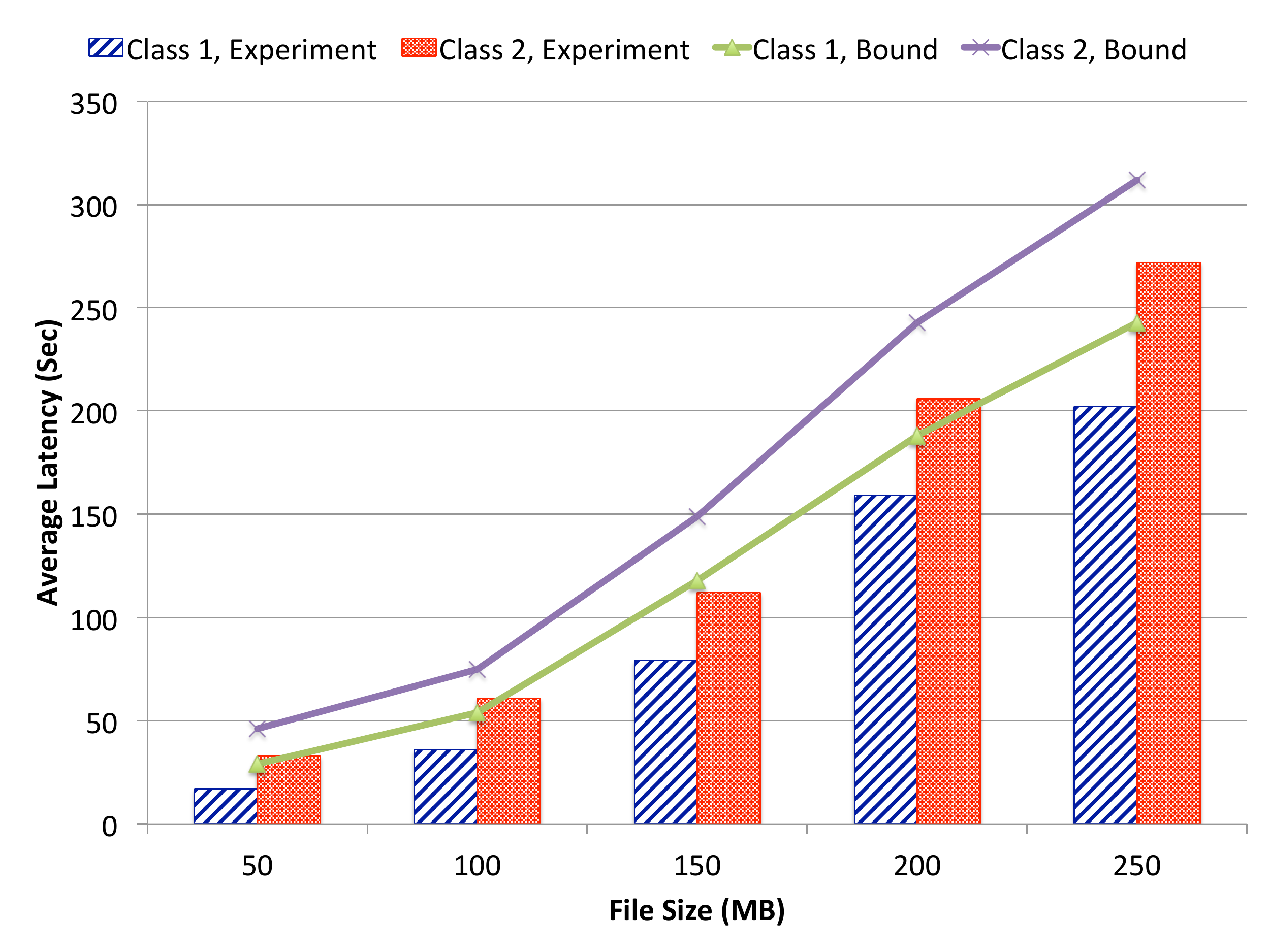}
\end{center}
\vspace{-0.2in}
\caption{\small Evaluation of different file sizes in our weighted queuing model for both class 1 and class 2 files with  $C_1=1$ and $C_2=0.4$. Aggregate rate for all file requests at 0.25/sec. Compared with class 2 files, our algorithm provides class 1 files relatively lower latency with heterogeneous file sizes. Latency increases as file size increases in both classes. Our analytic latency bound taking both network and queuing delay into account tightly follows actual service latency for both classes.}
\vspace{-0.1in}
\label{fig:size}
\end{minipage}
\end{figure*}
%\hspace{0.7cm}
%\begin{minipage}{0.46\textwidth}
\begin{figure}[!thbp]
\begin{center}
%\vspace{-0.5in}
\includegraphics[scale=0.33]{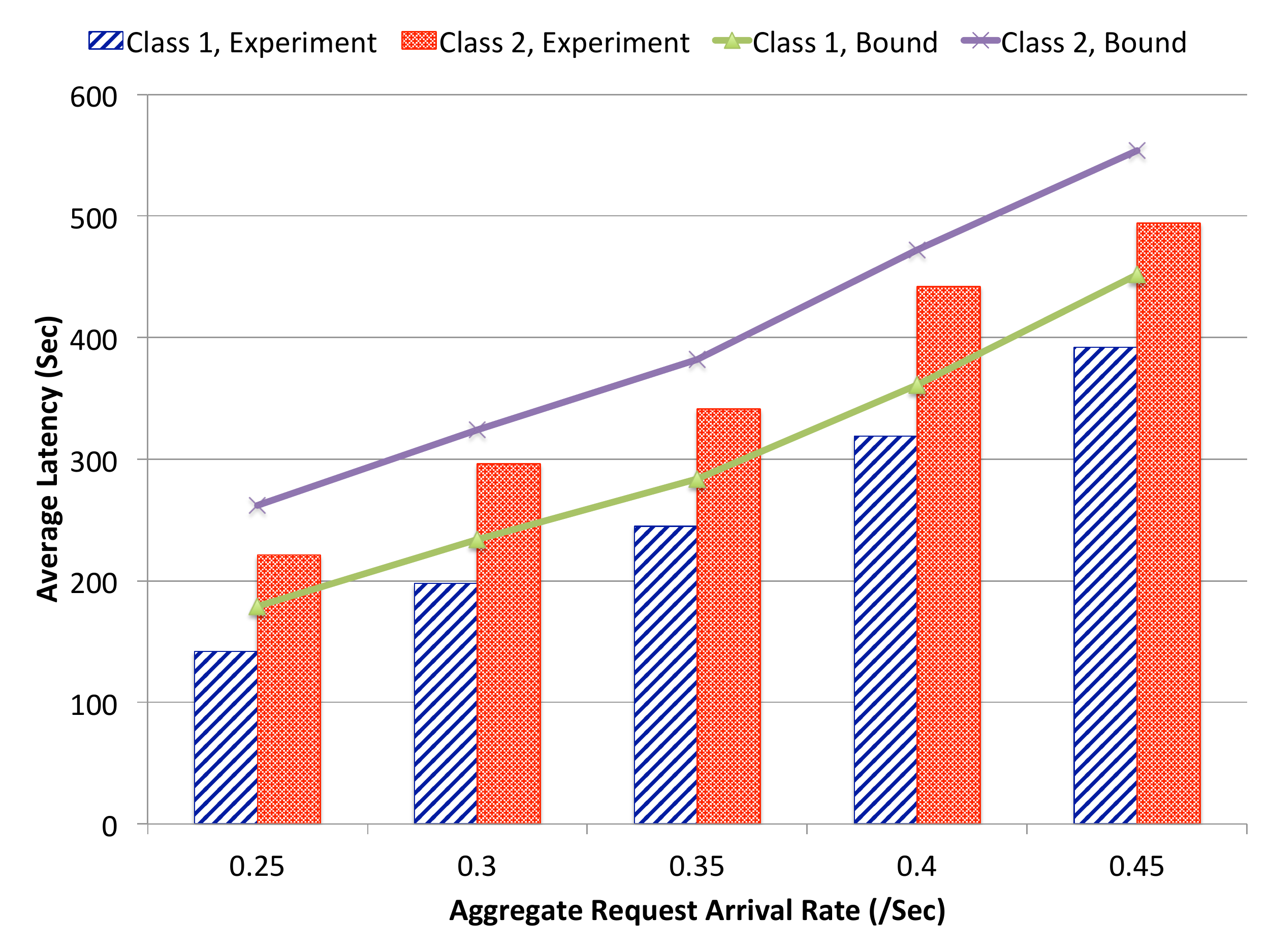}
\end{center}
\vspace{-0.2in}
\caption{\small Evaluation of latency performance for both classes of files with different request arrival rate in weighted queuing. File size 200M, with $C_1=1$ and $C_2=0.4$. Compared with class 2 files, our algorithm provides class 1 files relatively lower latency with heterogeneous request arrival rates. Latency increases as requests arrive more frequently for both classes. Our analytic latency bound taking both network and queuing delay into account tightly follows actual service latency for both classes. }
\vspace{-0.1in}
\label{fig:lambda}
%\end{minipage}
\end{figure}

%% file: Implep.tex
{\bf Validate algorithms and joint optimization.} In order to validate that Algorithm JLWO works for our system model (with weighted queuing model  for 2 classes of files for inter-rack traffic at the switch), we compare the average latency performance for the two file classes in our model in the cases of different access patterns. In this experiment, we have 1000 files for each class, all files have the same file size 200MB, and aggregate arrival rate for both classes from all racks is 0.25/sec. i.e., there is one request for each unique file, so the ratio of requests for the two classes are 1:1, and we set $C_1=1$ and $C_2=0.4$.  We are measuring the latency for the two classes of the following access patterns: 100\% concentration means 100\% of the requests for both classes of files concentrate on only one of the racks. Similarly, 80\% or 50\% concentration means 80\% or 50\% of the total requests of each file come from one rack, the rest of the requests spread uniformly among other racks.
Uniform access means that the requests for both classes are uniformly distributed across all racks.  We compare average latency for these requests for the two classes in each case of the access patterns with our weighted queuing model.

As shown in Fig ~\ref{fig:access}, experimental results indicate that our weighted queuing model can effectively mitigate the long latency due to congestion at one rack for both class 1 and class 2, as there is not a significant increase in average latency for both classes as the access pattern becomes more concentrated on one rack. Also we can see that class 1 has relatively lower latency than class 2 files since algorithm JLWO effectively assigned more bandwidth to requests for class 1 files since $C_1$ has a larger value which means class 1 files play a more important role in the optimization objective. It can also be observed that the difference in average latency for the two classes becomes less as the requests becomes more distributed, which is because our algorithm JLWO tentatively assigns more bandwidth to class 2 requests when the system is less congested than when it's highly congested, which matches our goal to minimize weighted average latency for both classes of files. We also calculated our analytic bound of average latency for the two classes when the above access patterns are applied. From the figure we can also see that our analytic bound is tight enough to follow the actual average latency for both class 1 and class 2 files.
%For example, when the request concentration level is 100\%, weighted queuing improves average latency by 32\%, and when concentration level is 80\%, the improvement is 27\%, as compared to the 24\% improvement provided by 50\% concentration and 21\% by the uniform distribution. We can see that this improvement in average latency increases as the requests become more concentrated, while in this case with our weighted queuing and the optimal chunk placement and retrieval scheduling, we see more weights allocated to the queues that have much heavier traffic than others. Optimal weight allocation allows us to utilize bandwidth more efficiently and reduce overall latency.

In order to further validate that the algorithms work for our weighted queuing models for the two classes, we  choose $r=1000$ files of size $100MB$ for each class, and aggregate arrival rate 0.25/sec for all file requests. the ratio of requests for the two classes are 1:1. We fix $C_1=1$ and vary $C_2$ from 0 to 2 and run Algorithm JLWO to generate the optimal solution for both file classes. The algorithms provide chunk placement, request scheduling and inter/intra-rack weight assignment ($W_{i,j,d}$ and $w_{i,j,d}$). From Fig. ~\ref{fig:sl-c2} we can see latency of class 2 files increases as $C_2$ decreases; i.e., when class 2 becomes less important. Also,  the average latency of class 1 requests decreases as $C_2$ decreases. This shows the expected result that when class 2 becomes more important, more weight is allocated to class 2, and since $C_2$ is always smaller than $C_1$, class 1 gets more bandwidth.  This also explains why when $C_1=C_2=1$ the average latency of the two classes are almost equal as shown in Fig. ~\ref{fig:sl-c2}. It can also be observed that when $C_2=0$ the system is only minimizing the latency of class 1 files, and thus class 2 files have a significantly large latency on average. The difference in latency between class 1 and class 2 become less when $C_2=2$ than when $C_2=0$ since in the former case algorithm JLWO is optimizing class 1 files as well, though at less importance with $C_1=1$.

{\bf Evaluate the performance of our solution} To demonstrate the effectiveness of our algorithms, we vary file size in the experiments from 50MB to 250MB.  In this experiment we still have 1000 files for each class,  aggregate arrival rate for both classes from all racks is 0.25/sec, so again the ratio of requests for the two classes are 1:1. We assume uniform random access, i.e., each file will be uniformly accessed from ten racks in the data center with a certain request arrival rate. Upload/download server selection is based on the algorithm output $S_i/\pi_{i,j,d}$, and inter/intra-rack bandwidth reserved according to output $W_{i,j,d}$ and $w_{i,j,d}$ from the optimization algorithm. Then we submit $r=1000$ requests for each class of files from the clients distributed among the racks. Results in Fig ~\ref{fig:size} show that our algorithm provides class 1 files relatively lower latency with heterogeneous file sizes compared to that of class 2 files, which means algorithm JLWO effectively assigned more bandwidth to requests for class 1 files since $C_1$ has a larger value which means class 1 files plays a more important role in the optimization objective. For instance, latency of class 1 files has a 30\% improvement on average over that of class 2 files for the 5 sample file sizes in this experiment.  Latency increases as requested file size increases for both classes when arrival rates are set to be the same. Since larger file size means longer service time, it increases queuing delay and thus average latency. We also observe that our analytic latency bound follows actual average service latency for both classes of files in this experiment. We note that the actual service latency involves other aspects of delay beyond queuing delay, and the results show that optimizing the metric of the proposed latency upper bound improves the actual latency with the weighted queuing models.

Similarly, we vary the aggregate arrival rate at each rack from 0.25/sec to 0.45/sec. This time we fix all file requests for file size 200MB. All other settings of this experiment are the same as the previous one whose result is shown in Fig ~\ref{fig:size}. In this experiment we also compare the latency of requests for the two file classes when we have different weights of their latency in optimization objective. (i.e., $C_1=1$ and $C_2=0.4$). We assume the same uniform access as before. For weighted queuing model, we use optimized server selection for upload and download for each file request, and optimal inter/intra-rack bandwidth reservation from Algorithm JLWO. Clients across the data center submit $r=1000$ requests for each file class with an aggregate arrival rate varying from 0.25/sec to 0.45/sec. From Fig ~\ref{fig:lambda} we can see that our algorithm provides class 1 files relatively lower latency with heterogeneous file sizes compared to that of class 2 files, which means algorithm JLWO effectively assigned more bandwidth to requests for class 1 files. For instance, latency of class 1 files has a 32\% improvement on average over that of class 2 files for the 5 sample aggregate arrival rate in this experiment.  Further,  as the arrival rate increases and there is more contention at the queues, this improvement becomes more significant. Thus our algorithm can mitigate traffic contention and reduce latency very efficiently. Also the average latency increases as the request arrival at each rack increases since waiting time increases with the workload for both classes.

%% file: single_dc_latency.bbl
\begin{thebibliography}{80}
%\vspace{-.1in}
\bibitem{Google}
E. Schurman and J. Brutlag, `` The user and business impact of server delays, additional bytes and http
chunking in web search," {\em O’Reilly Velocity Web performance and operations conference}, June 2009.


\bibitem{Lu:10}
Y. Lu, Q. Xie, G. Kliot, A. Geller, J. Larus, and A. Greenberg, ``Joinidle-queue: A novel load balancing algorithm for dynamically scalable web services," in Proc. {\em  IFIP Perforamnce}, 2010.

\bibitem{design}
Dell data center design, ``Data Center Design Considerations with 40GbE and 100GbE,", Aug 2013.

%\bibitem{DPR04}
%A.G. Dimakis, V. Prabhakaran, and K. Ramchandran, ``Distributed data storage in sensor networks using decentralized erasure codes," {\em Signals, Systems and Computers, 2004. Conference Record of the Thirty-Eighth Asilomar.}, 2004.

%\bibitem{AJX05}
%M.K. Aguilera, R. Janakiraman, L. Xu, ``Using Erasure Codes Efficiently for Storage in a Distributed System," {\em Proceedings of the 2005 International Conference on DSN}, pp. 336-345, 2005.


%\bibitem{HS07}
%H. Kameyam and Y. Sato, ``Erasure Codes with Small Overhead Factor and Their Distributed Storage Applications," {\em CISS '07. 41st Annual Conference}, 2007.

\bibitem{SH07}
A. Fallahi and E. Hossain, ``Distributed and energy-Aware MAC for differentiated services wireless packet networks: a general queuing analytical framework," {\em IEEE CS, CASS, ComSoc, IES, SPS}, 2007.

\bibitem{A98}
A.S. Alfa, ``Matrix-geometric solution of discrete time MAP/PH/1 priority queue," {\em Naval research logistics}, vol. 45, 00. 23-50, 1998.

\bibitem{TI10}
N.E. Taylor and Z.G. Ives, ``Reliable storage and querying for collaborative data sharing systems," {\em IEEE ICED Conference}, 2010.

\bibitem{KL98}
J.H. Kim and J.K. Lee, ``Performance of carrier sense multiple access with collision avoidance in wireless LANs," {\em Proc. IEEE IPDS.}, 1998.

\bibitem{ZA02}
E. Ziouva and T. Antoankopoulos, ``CSMA/CA Performance under high traffic conditions: throughput and delay analysis," {\em Computer Comm}, vol. 25, pp. 313-321, 2002.

\bibitem{SX}
S. Mochan and L. Xu, ``Quantifying Benefit and Cost of Erasure Code based File Systems." {\em Technical report available at $http://nisl.wayne.edu/Papers/Tech/cbefs.pdf$}, 2010.


\bibitem{WK}
H. Weatherspoon and J. D. Kubiatowicz, ``Erasure Coding vs. Replication: A Quantitative Comparison." {\em In Proceedings of the First IPTPS},2002

\bibitem{AG14}
C. Angllano, R. Gaeta and M. Grangetto, ``Exploiting Rateless Codes in Cloud Storage Systems," {\em IEEE Transactions on Parallel and Distributed Systems}, Pre-print 2014.

\bibitem{QFS13}
M. Ovsiannikov, S. Rus, D. Reeves, P. Sutter,S. Rao and J. Kelly, ``The quantcast file system," {\em Proceedings of the VLDB Endowment}, vol. 6, pp. 1092-1101, 2013.


\bibitem{MG1:12}
L. Huang, S. Pawar, H. Zhang and K. Ramchandran, ``Codes Can Reduce Queueing Delay in Data Centers," {\em Journals CORR}, vol. 1202.1359, 2012.

\bibitem{MDS-Queue}
N. Shah, K. Lee, and K. Ramachandran, ``The MDS queue: analyzing latency performance of erasure codes,"
{\em Information Theory (ISIT), 2014 IEEE International Symposium on}, July. 2014.


\bibitem{Makowski:89}
F. Baccelli, A. Makowski, and A. Shwartz, ``The fork-join queue and related systems with synchronization constraints: stochastic ordering and computable bounds,” {\em Advances in Applied Probability}, pp. 629–660, 1989.

\bibitem{Joshi:13}
G. Joshi, Y. Liu, and E. Soljanin, ``On the Delay-Storage Trade-off in Content Download from Coded Distributed Storage Systems," {\em arXiv:1305.3945v1}, May 2013.

\bibitem{DAB06}
B. Dekeris, T. Adomkus, A. Budnikas, ``Analysis of qos assurance using weighted fair queueing (WQF) scheduling discipline with low latency queue (LLQ)," {\em Information Technology Interfaces, 28th International Conference on}, 2006.

\bibitem{AT06}
M. Ashour, N. Le-Ngoc, ``Performance Analysis of Weighted Fair Queues with Variable Service Rates," {\em Digital Telecommunications, International Conference on}, 2006. ICDT '06.

\bibitem{ML10}
M.L. Ma, J.Y.B. Lee, ``," {\em Peer-to-Peer Computing (P2P),  IEEE Tenth International Conference on}, 2010.

\bibitem{XC07}
P. Xie,  J.H. Cui, ``An FEC-based Reliable Data Transport Protocol for Underwater Sensor Networks," {\em Computer Communications and Networks, Proceedings of 16th International Conference on}, 2007. ICCCN 2007.

\bibitem{YA10}
Y. Yang, K.M.M. Aung, E.K.K. Tong, C.H. Foh, ``Dynamic Load Balancing Multipathing in Data Center Ethernet," {\em Modeling, Analysis \& Simulation of Computer and Telecommunication Systems (MASCOTS), IEEE International Symposium on},2010.

\bibitem{LK13}
Bu. Lee, R. Kanagavelu, K.M.M. Aung, ``An efficient flow cache algorithm with improved fairness in Software-Defined Data Center Networks," {\em Cloud Networking (CloudNet), IEEE 2nd International Conference on}, 2013.


\bibitem{CS14}
S. Chen, Y. Sun, U.C. Kozat, L. Huang, P. Sinha, G. Liang, X. Liu and N.B. Shroff, `` When Queuing Meets Coding: Optimal-Latency Data Retrieving Scheme in Storage Clouds," {\em IEEE Infocom}, April 2014.

\bibitem{YW14}
O. N. C. Yilmaz, C. Wijting, P. Lunden, J. Hämäläinen, ``Optimized Mobile Connectivity for Bandwidth- Hungry, Delay-Tolerant Cloud Services toward 5G," {\em Wireless Communications Systems (ISWCS), 11th International Symposium on}, 2014.

\bibitem{NF12}
D. Niu, C. Feng and B. Li, ``Pricing cloud bandwidth reservations under demand uncertainty," {\em Proceedings of the 12th ACM SIGMETRICS/PERFORMANCE joint international conference on Measurement and Modeling of Computer Systems}, pp. 151-162, June 2012.

\bibitem{SP11}
S.Suganya and Dr.S.Palaniammal, ``A Well-organized Dynamic Bandwidth Allocation Algorithm for MANET," {International Journal of Computer Applications}, vol. 30(9), pp. 11-15, September 2011.
\bibitem{KT14}
A. Kumar, R. Tandon and T.C. Clancy, ``On the Latency of Erasure-Coded Cloud Storage Systems," {arXiv:1405.2833}, May 2014.

\bibitem{Yu-IFIP}
Y. Xiang, T. Lan, V. Aggarwal, and Y. R. Chen, ``Joint Latency and Cost Optimization for Erasure-coded
Data Center Storage," {\em Proc. IFIP Performance}, Oct. 2014 (available at arXiv:1404.4975 ).


\bibitem{MOSEK}
MOSEK, ``MOSEK: High performance software for large-scale LP, QP, SOCP, SDP and MIP," available online at {\em http://www.mosek.com/}.

\bibitem{hung}
Hungarian Algorithm, available online at {\em http://www.hungarianalgorithm.com}

\bibitem{OS:12}
D. Bertsimas and K. Natarajan, ``Tight bounds on Expected Order Statistics," {\em Probability in the Engineering and Informational Sciences}, 2006.


\bibitem{AY11}
A. Abdelkefi and J. Yuming, ``A Structural Analysis of Network Delay," {\em Ninth Annual CNSR}, 2011.

\bibitem{PT12}
F. Paganini, A. Tang, A. Ferragut and L.L.H. Andrew, ``Network Stability Under Alpha Fair Bandwidth Allocation With General File Size Distribution," {\em IEEE Transactions on Automatic Control}, 2012.

\bibitem{D11}
A.B. Downey, ``The structural cause of file size distributions," {\em Proceedings of Ninth International Symposium on MASCOTS}, 2011.


%\bibitem{LK13}
%G. Liang and U.C. Kozat, ``TOFEC: Achieving Optimal Throughput-Delay Trade-off of Cloud Storage Using Erasure Codes," {\em IEEE Infocom}, April 2014.

%\bibitem{SV14}
%V. Shah and G. Veciana, ``Performance Evaluation and Asymptotics for Content Delivery Networks," {\em IEEE Infocom}, April 2014.




%\bibitem{dummy}
%A.D. Luca and M. Bhide, ``Storage virtualization for dummies, Hitachi Data Systems Edition," {\em John and Wiley Publishing}, 2009.
%
%\bibitem{AmazonS3}
%Amazon S3, ``Amazon Simple Storage Service," {\em available online at http://aws.amazon.com/s3/}.
%
%\bibitem{Sathiamoorthy13}
%Sathiamoorthy, Maheswaran, et al. ``Xoring elephants: Novel erasure codes for big data.'' Proceedings of the 39th international conference on Very Large Data Bases. VLDB Endowment, 2013.
%
%\bibitem{Fikes10}
%Fikes, Andrew. ``Storage architecture and challenges.'' Talk at the Google Faculty Summit,{\em available online at http://bit.ly/nUylRW}, 2010.
%
%\bibitem{AJX05}
%M. K. Aquilera, HP Labs., P. Alto, R. Janakirama and L. Xu, ``Using erasure codes efficiently for storage in a distributed system," {\em Dependable Systems and Networks, 2005. DSN 2005. Proceedings. International Conference}, 2005.
%
%\bibitem{Dimakis}
%A. G. Dimakis, K. Ramchandran, Y. Wu, C. Suh, ``A Survey on Network Codes for Distributed Storage,'' arXiv:1004.4438, Apr. 2010













%\bibitem{RL05}
%R. Rosemark and W.C. Lee, ``Decentralizing query processing in sensor networks," {\em Proceedings of the second MobiQuitous: networking and services}, 2005



%\bibitem{RCCK12}
%R. Rojas-Cessa, L. Cai and T. Kijkanjanarat, ``Scheduling memory access on a distributed cloud storage network," {\em IEEE 21st annual WOCC}, 2012.



%\bibitem{CR09}
%S. Chen, K.R. Joshi and M.A. Hiltunem, ``Link Gradients: Predicting the Impact of Network Latency on Multi-Tier
%Applications," {\em Proc. IEEE INFOCOM}, 2009.




%\bibitem{LC02}
%Q. Lv, P. Cao, E. Cohen, K. Li and S. Shenker, ``Search and replication in unstructured peer-to-peer networks," {\em Proceedings of the 16th ICS}, 2002.

%\bibitem{LDT11}
%L. Huang, S. Pawar, H. Zhang, and K. Ramchandran, ``Codes can reduce queueing delay in data centers,” in Proc. IEEE ISIT 2012.

%\bibitem{DAB08}
%A. Duminuco, S. Antipolis, E.W. Biersack, ``Hierarchical Codes: How to Make Erasure Codes Attractive for Peer-to-Peer Storage Systems," {\em Peer-to-Peer Computing , Eighth International Conference}, 2008.



%\bibitem{LT10}
%H.Y. Lin, and W.G. Tzeng, ``A Secure Decentralized Erasure Code for Distributed Networked Storage," {\em Parallel and Distributed Systems, IEEE Transactions}, 2010.

%\bibitem{LWS09}
%W. Luo, Y. Wang and Z. Shen, ``On the impact of erasure coding parameters to the reliability of distributed brick storage systems," {\em Cyber-Enabled Distributed Computing and Knowledge Discovery, International Conference}, 2009.

%\bibitem{J06}
%J. Li, ``Adaptive Erasure Resilient Coding in Distributed Storage," {\em Multimedia and Expo, 2006 IEEE International Conference}, 2006.

%\bibitem{BSK11}
%K. V. Rashmi, N. Shah, and V. Kumar, ``Enabling node repair in any erasure code for distributed storage," {\em  Proceedings of IEEE ISIT}, 2011.

%\bibitem{WXHH06}
%X. Wang, Z. Xiao, J. Han and C. Han, ``Reliable Multicast Based on Erasure Resilient Codes over InfiniBand," {\em Communications and Networking in China, First International Conference}, 2006.



%\bibitem{CE04}
%P. Corbett, B. English, A. Goel, T. Grcanac, S. Kleiman, J. Leong and S. Sankar, ``Row-diagonal parity for double disk failure correction," {\em In Proceedings of the 3rd USENIX FAST'}, pp. 1-14, 2004.
%
%\bibitem{Calder:11}
%B. Calder, J. Wang, A. Ogus, N. Nilakantan, A. Skjolsvold, S. McKelvie, Y. Xu, S. Srivastav, J. Wu, H. Simitci, et al., `` Windows azure storage: A highly available cloud storage service with strong consistency," {\em In Proceedings of the Twenty-Third ACM SOSP}, pages 143--157, 2011.
%
%
%\bibitem{Khan:12}
%O. Khan, R. Burns, J. Plank, W. Pierce, and C. Huang, ``Rethinking erasure codes for cloud file systems: Minimizing I/O for recovery and degraded reads," {\em In Proceedings of FAST}, 2012.



%\bibitem{Ana:11}
%G. Ananthanarayanan, S. Agarwal, S. Kandula, A Greenberg, and I. Stoica, ``Scarlett: Coping with skewed content popularity in MapReduce," {\em Proceedings of ACM EuroSys}, 2011.

\bibitem{Bramson:10}
M. Bramson, Y. Lu, and B. Prabhakar, ``Randomized load balancing with general service time distributions," {\em Proceedings of ACM Sigmetrics}, 2010.


\bibitem{OS:12}
D. Bertsimas and K. Natarajan, ``Tight bounds on Expected Order Statistics," {\em Probability in the Engineering and Informational Sciences}, 2006.



%\bibitem{Boyd:05}
%S. Boyd and L. Vandenberghe, ``Convex Optimization," {\em Cambridge University Press}, 2005.
%
%\bibitem{DCP:12}
%L.T. Hoai An and P.D. Tao,``The DC (Difference of Convex Functions) Programming and DCA Revisited with DC Models of Real World Non-convex Optimization Problems," {\em Annals of Operations Research}, vol. 133, Issue 1-4, pp. 23-46, Jan 2005.

\bibitem{DC-Network}
Greenberg, A., Hamilton, J., Maltz, D. A., and Patel, P.,  ``The cost of a cloud: research problems in data center networks'', ACM SIGCOMM computer communication review, 39(1), 68-73, 2008.

\bibitem{Tahoe}
B. Warner, Z. Wilcox-O'Hearn, and R. Kinninmont, ``Tahoe-LAFS docs," {\em available online at https://tahoe-lafs.org/trac/tahoe-lafs}.

%\bibitem{MOSEK}
%MOSEK, ``MOSEK: High performance software for large-scale LP, QP, SOCP, SDP and MIP," available online at {\em http://www.mosek.com/}.
%
%\bibitem{Angell:02}
%T. Angell, ``The Farkas-Minkowski Theorem".  Lecture nodes available online at {\em www.math.udel.edu/$\sim$angell/Opt/farkas.pdf}, 2002.
%@article{dongarra2003linpack,
%  title={The LINPACK Benchmark: past, present and future},
%  author={},
%  journal={Concurrency and Computation: Practice and Experience},
%  volume={15},
%  number={9},
%  pages={803--820},
%  year={2003},
%  publisher={Wiley Online Library}
%}
\end{thebibliography}
